\documentclass[aps,prb,twocolumn,showpacs,citeautoscript]{revtex4-1}

\usepackage{bm}
\usepackage{amsmath, amsfonts}
\usepackage{graphicx}
\usepackage[colorlinks, linkcolor=blue, citecolor=blue, urlcolor=blue,
breaklinks]{hyperref}

\renewcommand{\vec}[1]{\bm{#1}}
\newcommand{\dotvec}[1]{\dot{\vec{#1}}}

\renewcommand{\Re}{\operatorname{Re}}

\begin{document}
\title{Dual approach to circuit quantization using loop charges}
\author{Jascha Ulrich}
\email{ulrich@physik.rwth-aachen.de}
\author{Fabian Hassler}
\affiliation{JARA-Institute for Quantum Information, RWTH Aachen University,
	D-52056 Aachen, Germany}
\date{May 2016}

\begin{abstract}
The conventional approach to circuit quantization is based on node fluxes and
traces the motion of node charges on the islands of the circuit. However, for
some devices, the relevant physics can be best described by the motion of
polarization charges over the  branches of the circuit that are in general
related to the node charges in a highly nonlocal way. Here, we present a
method, dual to the conventional approach, for quantizing planar circuits in
terms of loop charges. In this way, the polarization charges are directly
obtained as the differences of the two loop charges on the neighboring loops.
The loop charges trace the motion of fluxes through the circuit loops.  We
show that loop charges yield a simple description of the flux transport across
phase-slip junctions. We outline a  concrete construction of circuits based on
phase-slip junctions that are electromagnetically dual to arbitrary planar
Josephson junction circuits.  We argue that loop charges also yield a simple
description of the flux transport in conventional Josephson junctions shunted
by large impedances.  We show that a mixed circuit description in terms of
node fluxes and loop charges yields an insight into the flux
decompactification of a Josephson junction shunted by an inductor.  As an
application, we show that the fluxonium qubit is well approximated as a
phase-slip junction for the experimentally relevant parameters. Moreover, we
argue that the $0$-$\pi$ qubit is effectively the dual of a Majorana Josephson
junction.
\end{abstract}

\maketitle 

\section{Introduction}

Superconducting circuits offer the opportunity to study quantum mechanics on
mesoscopic scales unimpeded by dissipation.  The great flexibility in design
of the superconducting circuits has created the field of circuit quantum
electrodynamics where superconducting circuits are used as artificial atoms
featuring strongly enhanced light-matter coupling compared to standard cavity
QED.  Due to weak dissipation, such systems can be described
quantum-mechanically with an appropriate Hamiltonian.  Finding such a
Hamiltonian is the task of circuit quantization.  In recent years, there has
been a large interest in realizing purely reactive impedances, called
``superinductances'' $L$, with small parasitic capacitance $C$ such that the
characteristic impedance $Z=\sqrt{L/C}$ is much larger than the
superconducting resistance quantum $R_Q = h/4 e^2$~\cite{masluk:12}.  The large
impedance leads to a strong localization of charges with fluctuations below the
single Cooper-pair limit.  This fact makes these large inductances highly
relevant for qubits such as the $0$-$\pi$ qubit~\cite{kitaev:06a} or the
fluxonium~\cite{manucharyan:09} with strongly reduced sensitivity to external
charge fluctuations.  The suppression of charge fluctuations below the single
Cooper-pair limit is also relevant for phase slip junctions.  Considering the
transport of quantized fluxoids as duals of the quantized electron
charge~\cite{tinkham}, phase-slip junctions can be understood as electric duals
of conventional Josephson junctions with a nonlinear, $2e$-periodic
voltage-charge relation $V(Q)$~\cite{mooij:06}.  Recently, there has been much
progress both in the theoretical understanding~\cite{arutyunov:08} and the
experimental realization~\cite{astafiev:12,peltonen:13,belkin:11,belkin:15} of
phase slip junctions using superconducting nanowires.  Large characteristic
impedances also imply strongly enhanced electric fields in waveguides, allowing
an enhanced coupling to qubits like the transmon or efficient nano-mechanical
coupling to nanostructures~\cite{fink:16,samkharadze:16}.

The localization of charge in circuits with large impedances suggests a
description in terms of the polarization charges on the circuit elements which
remain close to being good quantum variables due to their slow dynamics.  The
conventional approach to circuit quantization in terms of node fluxes,
however, works  with the charges on the islands, which are related to the
polarization charge in a highly nonlocal way~\cite{yurke:84,devoret:96}. While
the node-flux formalism is well-suited for the description of the fast charge
transport in superconducting devices with low impedances and localized fluxes,
it must be considered ill-suited for the description of fast flux transport
with localized charges in large-impedance environments. In particular, the
nonlinear capacitive behavior of phase-slip junctions cannot be modeled in a
straightforward way using node fluxes.

In view of the growing interest in superinductances and phase-slip junctions
in the large-impedance setting, we provide here a dual approach to
circuit quantization in terms of loop charges. As we will show, it yields a
simple description of planar circuits involving phase-slip junctions in the
same way as the use of node fluxes yields a simple description of circuits
involving Josephson junctions.  Loop charges are the time-integrated currents
circulating in the loops of a planar circuit and their canonical momenta are
the physical fluxes within the loops.  While in the node flux formulation terms
in the Hamiltonian relate to the transport of the physical charges on the
islands, the loop charge formulation describes the transport of the physical
fluxes within the loops~\cite{ivanov:01b,friedman:02}.  Therefore, the
formalism presented here will be most useful for problems for which it  is more
natural to think about the transport of fluxes rather than about the transport
of Cooper pairs.

Loop currents as independent current degrees of freedom were already
considered by Maxwell~\cite{maxwell:92} and are frequently used in mesh
analysis of electrical engineering.  However, due to the typically large
number of dissipative components in  electrical network, systematic Lagrangian
formulations have received only limited attention~\cite{shragowitz:88,
kwatny:82, chua:74, macfarlane:69, weiss:97, massimo:80} and are not tailored
specifically  to the problem of circuit quantization.  On the other hand, in
the superconducting community, the loop charge formulation appears to be
largely unknown.  Charge degrees of freedom akin to loop charges have
previously been introduced through explicit analysis of the Kirchhoff current
law~\cite{bakhalov:89, hermon:96, haviland:96, homfeld:11, vogt:15}.  An
explicit analysis of the Kirchhoff current law can be avoided by using matrix
representations of the circuit topology~\cite{burkard:04,burkard:05} at the
expense that  the Lagrangian cannot be read off straightforwardly from the
circuit graph.

In contrast, here we are interested in presenting a formulation that makes
circuit quantization straightforward in the sense that the Lagrangian can be
obtained immediately from the circuit graph using a set of simple rules.  In
Sec.~\ref{sec:circuit_quantization_node_fluxes}, we give a brief introduction
to the node flux formulation, including a more extensive discussion of its
problems with the description of phase-slip junctions.  In
Sec.~\ref{sec:circuit_quantization_loop_charges}, we introduce the new loop
charge formulation.  We provide simple rules for the construction of the
Lagrangian of a lumped element circuit and discuss the Legendre transform to
the Hamiltonian formulation.  We also discuss how to handle offset charges,
external fluxes, and voltage or current sources.  In Sec.~\ref{sec:duality}, we
discuss the duality between the node flux and the loop charge formulation.  In
Sec.~\ref{sec:duality_passive}, we consider passive duality transformations
where the same system is described using different variables and explicitly
construct the transformation from the node flux to the loop charge
representation of a given circuit.  This section may be skipped on first
reading since in practice it is sufficient and much easier to use the rules
given in Sec.~\ref{sec:circuit_quantization_loop_charges} for the construction
of the loop charge Lagrangian.  In Sec.~\ref{sec:duality_active}, we consider
active duality transformations which yield new circuits electromagnetically
dual to a given circuit.  We show how to construct electromagnetic duals of
arbitrary circuits using the loop charge formulation.  In
Sec.~\ref{sec:dissipation}, we discuss how to introduce dissipation in circuits
described by loop charges.  In Sec.~\ref{sec:mixed_formulation}, we extend the
formalism to mixed circuit descriptions where part of the circuit is described
in terms of node fluxes and some other part in terms of loop charges.  This
leads to additional insights regarding the flux decompactification of
inductively shunted Josephson junctions.\cite{koch:09} Finally, in
Sec.~\ref{sec:applications}, we discuss examples of the loop charge description
for the fluxonium and the $0$-$\pi$ qubit.  We show that for large inductances
the fluxonium qubit can be well approximated as a nonlinear capacitor and the
$0$-$\pi$ qubit effectively becomes the dual of a Majorana Josephson junction.
We finish with a short discussion of our results.

As a last point, let us, for the convenience of the reader, briefly comment on
the conventions and the terminology that we will use in this paper.  We will
represent a circuit as a directed graph which we will occasionally also refer
to as the (electrical) network.  Following conventions from electrical
engineering, we will also use the term branches when referring to the edges of
the circuit and the word node when referring to the vertices.  In contrast, we
will simply refer to the loops of the circuits as loops and refrain from using
the word meshes.  Throughout this work, $\phi$ will denote fluxes in terms of
which the superconducting phase differences are given by $ 2\pi \phi/\Phi_Q$
with the superconducting flux quantum $\Phi_Q = h/2e$.

\section{Circuit quantization using node fluxes or loop charges}\label{sec:circuit_quantization}

In the lumped element approximation, an electrical circuit is described as a
graph where each branch represents a two-terminal electrical element such as a
capacitor, an inductor, a voltage source, and so forth.  In order to
consistently keep track of the orientations, we assign an orientation to each
branch of the graph which specifies the direction in which a positive current
flows and the direction of a positive voltage drop.  The lumped element
approximation yields a simplified circuit description that is valid as long as
the propagation time of electromagnetic waves between the circuit elements is
negligible, i.e., the circuit dimensions are much smaller than the wave-length
of electromagnetic radiation at the frequencies of interest.  While in the
general case, characterizing the circuit requires the calculation of the
microscopic electric and magnetic fields within the circuit from Maxwell's
equations, within the lumped element approximation, it is sufficient to know
the voltage drops $V^\text{br}_b$ across and the currents $I^\text{br}_b$ along
each branch $b$ of the network.  The equations governing the behavior of the
voltages $V^\text{br}_b$ and the currents $I^\text{br}_b$ are the Kirchhoff
circuit laws and the element-dependent constitutive laws relate $V^\text{br}_b$
and $I^\text{br}_b$.

It is convenient to work exclusively with independent voltages $\vec{V}$ or
currents $\vec{I}$ which determine all the voltage drops
$\vec{V}^\text{br}(\vec{V})$ and current flows $\vec{I}^\text{br}(\vec{I})$
within the circuit in such a way that either the Kirchhoff voltage law or the
current law is automatically fulfilled.  The dynamics of the voltages $\vec V$
or currents $\vec I$ is governed by differential equations obtained after
applying the remaining Kirchhoff law together with the constitutive laws.  The
constitutive laws are most easily stated in terms of branch fluxes
$\vec{\phi}^\text{br}$ and branch charges $\vec{q}^\text{br}$ defined as
\begin{align}
\vec{\phi}^\text{br}(t) &= \int_{-\infty}^t \!d t' \, \vec{V}^\text{br}(t), 
\label{eq:phi_br}\\
\vec{q}^\text{br}(t) &= \int_{-\infty}^t \!d t' \, \vec{I}^\text{br}(t),
\label{eq:q_br}
\end{align}
where $\vec{V}^\text{br}$ and $\vec{I}^\text{br}$ are the vectors of branch
voltages and currents, respectively.  For a capacitor on branch $b$,
$q^\text{br}_b$ can be interpreted as the (polarization) charge on one of the
capacitor plates\cite{Note1} and the constitutive law assumes the form
\begin{align} \label{eq:constitutive_law_cap}
V_b^\text{br} = f_{V,b}(q^\text{br}_b),
\end{align}
where the voltage is given by $f_V(q) = q/C$ for an ideal capacitor $C$. For a
phase-slip junction, on the other hand, the function $f_V(q)$ is periodic with
period $2e$. In the simplest model, we obtain the expression $f_V(q) = V_c
\sin(\pi q/e)$, with $V_c$ the critical voltage.

For inductors, Faraday's law yields an interpretation of $\phi_b^\text{br}(t)$
as the flux threading the inductor and the constitutive law takes the form
\begin{align}\label{eq:constitutive_law_ind}
I_b^\text{br} = f_{I,b}(\phi_b^\text{br}),
\end{align}
with $f_I(\phi) = \phi/L$ for an ideal inductance $L$.  The constitutive
relations~\eqref{eq:constitutive_law_cap} and~\eqref{eq:constitutive_law_ind}
suggest that in general, it will be most convenient to work with independent
fluxes $\vec \phi$ or charges $\vec Q$ that are the time-integrated voltages
$\vec V$ or currents $\vec I$ defined in a way analogous to
Eq.~\eqref{eq:phi_br},~\eqref{eq:q_br} such that $\dotvec{\phi} = \vec V$ or
$\dotvec{Q} = \vec I$.  For circuit quantization, we are then interested in
finding a Lagrangian $\mathcal{L}(\vec \phi, \dotvec{\phi})$ or
$\mathcal{L}(\vec Q, \dotvec Q)$ such that its equations of motion reproduce
the differential equations originating from the remaining Kirchhoff law.

The choice between a flux-based or a charge-based approach is restricted by
two considerations.  The first restriction comes from circuit quantization.
For circuit quantization, we require the circuit Lagrangian $\mathcal{L}(\vec
x,\dotvec{x})$ for the degrees of freedom $x_i$ to be of the standard form
$\mathcal{L} = T(\dotvec{x}) - U(\vec x)$ known from classical mechanics,
where $T$ is a quadratic form corresponding to a kinetic energy term and $U$
is a potential energy term.  The other restriction comes from the constitutive
laws.  For example, the constitutive relation~\eqref{eq:constitutive_law_cap}
shows that the charge $q_b^\text{br}$ may be a convenient degree of freedom
for the description of a capacitor since it determines both the current
$I_b^\text{br} = \dot q_b^\text{br}$ and the voltage $V_b^\text{br}$ through
relation~\eqref{eq:constitutive_law_cap}.  Similarly, the flux
$\phi_b^\text{br}$ may be a convenient degree of freedom for the description
of an inductor since it determines the voltage $V_b^\text{br} = \dot
\phi_b^\text{br}$ and the current through
relation~\eqref{eq:constitutive_law_ind}.

We will start by reviewing the flux-based formulation in terms of node
fluxes~\cite{devoret:96} and then introduce the new charge-based formulation
in terms of loop charges.

\begin{figure}[tb]
\centering
\includegraphics{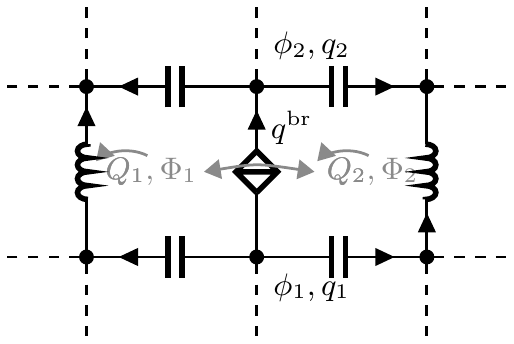}%
\caption{Example network with the loop charges $Q_1$ and $Q_2$, which are the
	time-integrated currents circulating in the loops in the specified
	orientation, and their conjugate momenta $\Phi_1$ and $\Phi_2$, which
	are the fluxes in the respective loops.  For comparison, we also
	indicate node fluxes $\phi_1$ and $\phi_2$ at two nodes (shown as
	dots) of the network together with their conjugate momenta $q_1$ and
	$q_2$ which are the charges on the islands.  For general networks the
	physical charge across a branch is related to the charges on the
	islands in a highly nonlocal way.  In contrast, it is easy to see that
	using the loop charges $Q_1$ and $Q_2$, we obtain the local expression
	$q_b^\text{br}= Q_1 - Q_2$ for the polarization charge across the
	phase-slip junction (diamond) taking their respective orientations
	into account.  We have also indicated the transverse flux flow through
	the phase-slip junction (gray double-headed arrow).  In contrast to a
	normal capacitor, in a phase-slip junction, the flow of flux is
	quantized in units of the superconducting flux quantum $\Phi_Q$.  This
	expresses the duality to a Josephson junction which features
	longitudinal charge-transport in the direction of the element which is
	quantized in units of the Cooper-pair charge $2e$.}
\label{fig:loop_charge_conservation}%
\end{figure}

\subsection{Node flux representation}\label{sec:circuit_quantization_node_fluxes}

The Kirchhoff voltage law states that the ``vector field'' $\vec \phi^\text{br}$
is conservative.  Therefore the Kirchhoff voltage law can automatically be
satisfied provided the fluxes $\vec \phi^\text{br}$ are represented via the
``gradient'' of a potential.  In the discrete graph setting, the potential is
given by the node fluxes $\phi_n$ that are placed on each node $n$ of the
circuit.  For a branch $b$ directed from node $n$ to node $n'$, the branch
flux $\phi^\text{br}_b$ is obtained as the discrete gradient $\phi^\text{br}_b
= \phi_{n} - \phi_{n'}$ of the node fluxes (along $b$).  In this way, the node
fluxes determine all the voltage drops over the branches of the circuit.
Since the physical voltages depend only on differences of node fluxes, we may
arbitrarily set the flux of one of the nodes (called the ground node) to zero.
The voltage $\dot \phi_n$ associated with a node flux can then be interpreted
as a voltage relative to ground.

The Kirchhoff current law is implemented through the equations of motion of a
Lagrangian $\mathcal{L}(\vec \phi, \dotvec{\phi})$ which is constructed as
follows.  Each inductive element at a branch $b$ adds the term
$-U(\phi_b^\text{br})$ to the Lagrangian, where
\begin{align}\label{eq:pot_energy_ind}
U(\phi_b^\text{br}) = \int_{0}^{\phi_b^\text{br}(t)} \!d \phi\, f_{I,b}(\phi)  
\end{align}
is simply the magnetic field energy as can be easily verified by integrating
the power $V_b^\text{br}(t) I_b^\text{br}(t)$ over time and using the
relation~\eqref{eq:constitutive_law_ind}.  Similarly, each capacitive element
with capacitance $C$ adds a term $C \dot \phi_b^2/2$ which is just the
electric field energy.

The equations of motion with respect to a node flux $\phi_n$ are given by the
Euler-Lagrange equations
\begin{align}\label{eq:euler_lagrange_node_flux}
\frac{d}{dt} \frac{\partial \mathcal{L}}{\partial \dot \phi_n} - 
\frac{\partial \mathcal{L}}{\partial \phi_n} = 0.
\end{align}
Let us consider a branch directed from a node $n'$ towards a node $n$ such
that $\phi_b^\text{br} = \phi_{n'} - \phi_{n}$.  For inductive branches, we
obtain a term $-I_b^\text{br} = -f_{I,b}(\phi_b^\text{br})$ to the current
balance while for capacitive branches, we obtain a term $-C \ddot
\phi_b^\text{br}$.  In both cases, this is just the current flowing away from
node $n$ through branch $b$.  For the opposite orientation
$\phi_{b'}^\text{br} = \phi_{n} - \phi_{n'}$, we would obtain
$I_{b'}^\text{br} = f_{I,b'}(\phi_{b'}^\text{br})$ and $C \ddot
\phi_{b'}^\text{br}$.  In both cases, we therefore obtain the current flowing
away from node $n$.  We conclude that the equations of motion for the node
flux $\phi_n$ reproduce the Kirchhoff current law at node $n$.  The formalism
can straightforwardly be extended to include electromotive forces due to
external magnetic fields, see Ref.~\onlinecite{devoret:96}.

The form of the constitutive relation~\eqref{eq:constitutive_law_ind}
indicates that the node flux representation is well-suited for the description
of nonlinear inductances.  The knowledge of the branch flux $\phi_b^\text{br}$
over an inductance readily gives access to the voltage and the current through
Eq.~\eqref{eq:constitutive_law_ind}.  Moreover, the
terms~\eqref{eq:pot_energy_ind} added to the Lagrangian can simply be
interpreted as (possibly nonlinear) potential energy terms which pose no
problem for circuit quantization.

In contrast, the node flux formulation cannot be used for the description of
nonlinear capacitors.  The constitutive
relation~\eqref{eq:constitutive_law_cap} shows that the natural variable for a
capacitor is the branch charge $q_b^\text{br}$ rather than the branch flux
$\phi_b^\text{br}$.  Determining the current flow through the capacitor solely
from the knowledge of $\phi_b^\text{br}$ is generally impossible.  Although
for invertible $f_{V,b}$, we may in principle obtain $I_b^\text{br} =
\ddot{\phi}_b^\text{br}/f_{V,b}'[f_{V,b}^{-1}(\dot \phi_b^\text{br})]$,
generating this term through the equations of motion requires adding a term of
the form $\int_0^{\dot \phi_b^\text{br}} \!d V \, f^{-1}_{V,b}(V)$ to the
Lagrangian.  This will only lead to a quadratic kinetic energy term $C \dot
\phi_b^2/2$ when considering a linear capacitor $C$.  In contrast, a circuit
containing a nonlinear capacitance cannot readily be quantized when described
in terms of node fluxes $\phi_n$.  To that end, we need a charge-based
description which we will describe in details in the next section.

\subsection{Loop charge representation}
\label{sec:circuit_quantization_loop_charges}

While the idea of representing the `vector field' $\vec \phi^\text{br}$ by a
`scalar potential' $\vec \phi$ in order to guarantee the Kirchhoff voltage law
is rather natural, it may be less obvious how to define charge degrees of
freedom which automatically guarantee current conservation.  For a planar
graph that is effectively two-dimensional such that it can be drawn on a sheet
of paper without crossing lines, the correct degrees of freedom for that
purpose are the loop charges $Q_l$.  They are the time-integrated loop
currents circulating within every loop $l$ of the network that does not have
any inner loops, c.f.~Fig.~\ref{fig:loop_charge_conservation}.  We give an
orientation to the loop charges by specifying the orientation of a positive
current flow.  This orientation is in principle arbitrary but the simplest
rules emerge for a consistent choice of orientation.  In the current paper, we
choose the orientation of all loop currents to be counter-clockwise.

Similar to the node fluxes, the loop charges are unphysical degrees of freedom
in the sense that they generally do not correspond directly to a physical
charge on a branch of the network.  For example, by simple inspection of
Fig.~\ref{fig:loop_charge_conservation}, we observe that the polarization
charge $q^\text{br}_b$ of the phase-slip junction (diamond) on the branch $b$
in the specified direction is given by the difference $q^\text{br}_b = Q_1 -
Q_2$ of the loop charges with their indicated orientations; here, the loop
charge $Q_1$ ($Q_2$) enters with a plus (minus) sign as its orientation is
along (opposite) to that of $q^\text{br}_b$.  While in the node flux
formulation, we obtain the physical flux across every branch as the difference
of node fluxes on neighboring nodes, in the loop charge formulation, we obtain
in this way the physical (polarization) charge across every branch as the
difference of loop charges in neighboring loops.  By formally placing a loop
charge $Q_0 = 0$ at the exterior of the circuit, this statement also remains
correct for finite circuits with a boundary.
\begin{figure}[tb]
\centering
\includegraphics{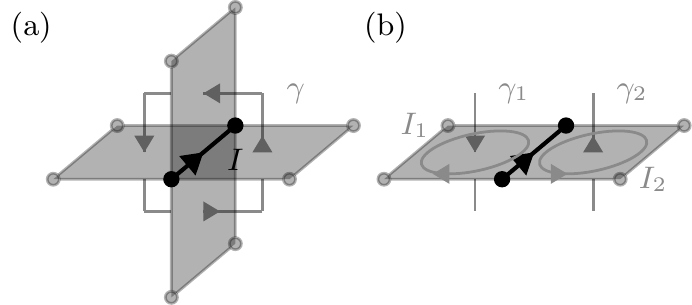}%
\caption{Motivation of loop currents from Maxwell's equations for a lumped
	element circuit, represented in terms of its nodes (dots) and faces
	(light filled rectangles).  In (a) a general network is shown with a
	current $I$  (dark arrow)  along  a directed branch of the circuit.
	The contour  $\gamma$ (light arrows) encircles the current $I$.
	According to Maxwell's equations, current conservation is guaranteed
	when the current $I$ running through the branch is obtained from the
	circulation of the magnetic field around it, $I \propto \oint_{\gamma}
	d\vec s \cdot \vec B$.  The total current $I$ can be decomposed into a
	sum of currents $I_l \propto \int_{\gamma_i} d \vec s \cdot \vec B$,
	where each part $\gamma_l$ of the contour is associated with a
	specific face of the circuit graph that is pierced by the contour. The
	currents $I_l$ have a direct interpretation in terms of the currents
	circulating around the pierced loops.  This is particularly easy to
	see for a planar circuit depicted in (b) as we may close the contour
	integral at infinity.  As a result, we obtain $I = (-I_1) - I_2$, in
	line with the interpretation of the currents $I_l$ as currents
	circulating in loop $l$.  Viewing the circuit from above, we recover
	Fig.~\ref{fig:loop_charge_conservation}.}%
\label{fig:current_conservation}%
\end{figure}

The loop charge construction can also be justified directly from Maxwell's
equations.  According to Maxwell's equations, current conservation (in a
stationary situation) is guaranteed when the current $I$ flowing through some
area bounded by a contour $\gamma$ is obtained from the circulation of the
magnetic field according to $I \propto \oint_{\gamma} d\vec s \cdot \vec B$.
For each branch $b$ of the network, we can decompose the current $I_b$ into a
sum of currents $I_l \propto \int_{\gamma_l} d \vec s \cdot \vec B$, where
each part $\gamma_l$ of the contour is associated with a specific face of the
circuit that is pierced by the contour, see
Fig.~\ref{fig:current_conservation}(a).  The current $I_l$ can be interpreted
as the loop current within the pierced loop $l$, see
Fig.~\ref{fig:current_conservation}(b).  The loop charge $Q_l$ is then simply
related to the current $I_l$ as $\dot Q_l = I_l$.  The above considerations
also show that we will generally only obtain the current from the difference
of precisely two loop charges when the circuit is planar, i.e., effectively
two-dimensional~\cite{thulasiraman}.  We show in
App.~\ref{app:math} that the loop charge description is indeed limited to
planar circuits.  

\begin{figure}[tb]
\centering
\includegraphics{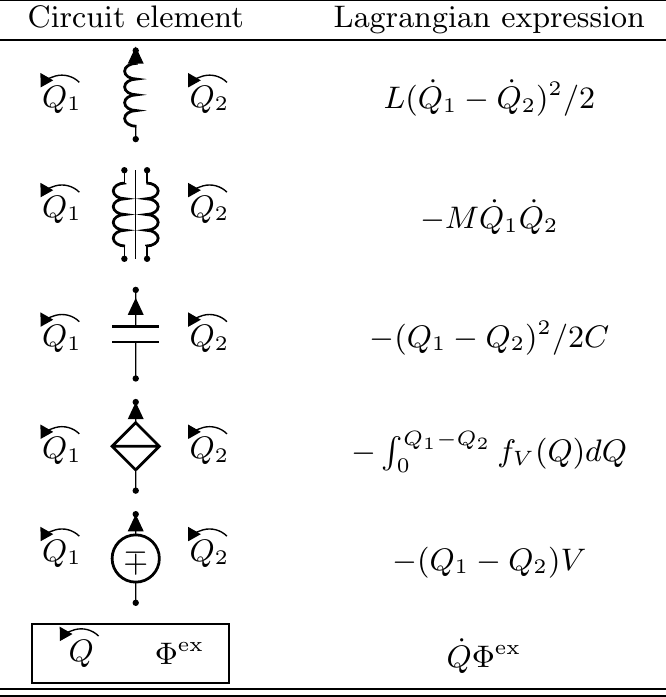}
\caption{The left column depicts various circuit elements [inductor $L$,
  capacitor $C$, mutual inductance $M$, general capacitance with
  voltage-charge relation $V= f_V(Q)$, voltage source $V$, and external flux
  $\Phi^\text{ex}$] with their corresponding expression in the Lagrangian
  (right column).  In a planar graph, each of the circuit elements is part of
  two loops with loop charges $Q_1$ and $Q_2$ which are indicated along with
  their respective orientation for completeness.  The simplest representation
  of a phase-slip junction amounts to choosing $f_V(Q) = (\pi E_S/e) \sin(\pi
  Q/e)$, where $E_S/\hbar$ is the phase-slip rate.  This corresponds to a term
  $E_S \cos[\pi(Q_1 - Q_2)/e]$ in the
  Lagrangian.\label{fig:loop_charge_summary}}
\end{figure}

Having identified the loop charges $\vec Q$ as variables guaranteeing current
conservation, we are left with the task of defining a Lagrangian whose
equations of motion guarantee the Kirchhoff voltage law.  The construction of
this Lagrangian is analogous to the construction of the Lagrangian for the
node fluxes.  Specifically, each capacitive element adds a term
$-U(q_b^\text{br})$ to the Lagrangian, where
\begin{align}
U(q_b^\text{br}) = \int_0^{q_b^\text{br}(t)} \!d q \, f_{V,b}(q) 
\end{align}
is just the electric energy stored in the capacitor.  Specifically, for the
simplest model $f_{V}(Q) = V_c \sin(\pi Q/e)$ of a phase-slip
junction,  we obtain the term (up to a constant)
\begin{align}
U(q_b^\text{br}) = -E_S \cos(\pi q_b^\text{br}/e)
\end{align}
with the characteristic energy $E_S = e V_c/\pi $.  For each linear inductor
$L$, we add a kinetic term of the form $L (\dot q_b^{\text{br}})^2/2$.  In
this way, the equations of motion with respect to a loop charge $Q_l$ yield
the balance of voltage drops obtained from a counter-clockwise traversal of
the loop $l$.  The relevant terms that have to be added to the Lagrangian are
summarized for different components in Fig.~\ref{fig:loop_charge_summary}.
Since Josephson junctions are nonlinear inductors, they cannot be directly
described using the loop charge formulation.  We will introduce a way to
obtain a charge-based descriptions of Josephson junctions in
Sec.~\ref{sec:applications} (see also the comments in
Sec.~\ref{sec:duality_passive}).

A Hamiltonian description requires the introduction of canonical momenta
\begin{align}\label{eq:loop_charge_canonical_momentum}
\Phi_l = \partial \mathcal{L}/\partial{\dot Q_l}.
\end{align}
Each $\Phi_l$ can be interpreted as the loop flux in the loop $l$ of the
circuit.  If the relation~\eqref{eq:loop_charge_canonical_momentum} between
the loop fluxes $\vec \Phi$ and the loop charges $\vec Q$ is invertible, we
can perform the Legendre transformation
\begin{align}\label{eq:legendre_transformation}
H = \vec \Phi \cdot \dotvec{Q} - \mathcal{L}(\vec Q, \dotvec{Q}) 
\end{align}
and obtain the circuit Hamiltonian which can be readily quantized through the
introduction of canonical commutation relations $[\Phi_{j}, Q_k] = \delta_{jk}
i \hbar$.

It may happen that the relation~\eqref{eq:loop_charge_canonical_momentum}
between the loop charges $\vec Q$ and the conjugate momenta $\vec \Phi$ is not
invertible.  This indicates that not all loop currents are dynamical degrees
of freedom.  A simple example for this is an inductor $L$ with two parallel
capacitances $C_1$ and $C_2$ to the left and the right.  Denoting the loop
charges in the two loops by $Q_1$ and $Q_2$, the corresponding Lagrangian
reads $\mathcal{L} = L (\dot Q_1 - \dot Q_2)^2/2 - Q_1^2/2C_1 - Q_2^2/2C_2$.
Introducing $Q = Q_1 - Q_2$ and $Q' = (Q_1+Q_2)/2$, it is obvious that the
state of the system depends only on the current $\dot Q$ through the inductor
and not on the currents through the capacitive branches.  As a consequence,
the Lagrangian does not depend on $\dot Q'$ which gives the constraint
$\partial \mathcal{L}/\partial \dot Q' = 0 = \Phi'$ for the momentum $\Phi'$
conjugate to $Q'$ which cannot be solved for $\dot Q'$.  However, the fact
that the Lagrangian does not depend on $\dot Q'$ also means that the
Euler-Lagrange equations for $Q'$ are purely algebraic equations (constraints)
which can be solved immediately.  Resolving the constraint for $Q'$ and
reinserting  the solution into the Lagrangian yields the regular Lagrangian
$\mathcal{L} = L \dot Q^2/2 - Q^2/2 (C_1 + C_2)$.  Resolving all constraints
in such a way in general leads to a reduced Lagrangian involving only
dynamical degrees of freedom such that the Legendre
transformation~\eqref{eq:legendre_transformation} and quantization can be
performed.

Superconducting circuits with Josephson junctions or phase-slip junctions may
involve transport of strictly quantized charges or fluxes through the circuit.
The former situation occurs when a superconducting island is connected to the
rest of the network only by capacitors and Josephson junctions.  The isolation
of the island demands that the node charge $q_n$ of the island is quantized in
units of $2e$ which corresponds to a $\Phi_Q$-periodicity of the wavefunction
in terms of the node flux $\phi_n$.  The latter situation occurs if a loop $l$
involves only inductors and phase-slip junctions. In this case the flux
$\Phi_l$ in the loop is quantized in units of $\Phi_Q$ corresponding to a
$2e$-periodicity of the wavefunction with respect to the corresponding loop
charge $Q_l$.  

Instead of focusing on the circuit to identify islands with integer node
charges (in units of $2e$) or loops with integer loop fluxes (in units of
$\Phi_Q$) to determine the appropriate boundary conditions for the
quantization of the fluxes or charges, we may also determine the appropriate
choice of boundary conditions by looking at the symmetries of the Hamiltonian.
The quantization of fluxes or charges is due to the periodicity of the
underlying potentials.  If one ignores the periodicity considerations of the
wavefunction as described above and works with node fluxes $\vec \phi$ or loop
charges $\vec Q$ defined on the entire real axis, the periodicity leads to the
existence of conserved quantities which correspond to Bloch quasi-momenta.  A
specific choice of Bloch momentum then corresponds to a choice of initial
condition.  Due to the relations \eqref{eq:phi_br} and \eqref{eq:q_br}, our
inital condition for $t\to-\infty$ corresponds to a charge- and flux-less
state and thus all the Bloch momenta vanish (implying periodic
wave-functions).  The two approaches are therefore equivalent and one may
choose whatever method seems more convenient.  The symmetry-based perspective
will be particularly useful in the mixed formulation to be discussed in
Sec.~\ref{sec:mixed_formulation}.

A typical lumped element circuit does not just involve passive elements like
capacitors and inductors, but also involves active elements like voltage and
current sources.  It will also feature electromotive forces due to
time-varying fluxes or offset charges on some island of the network.  Voltage
sources generating a voltage drop $V_i^\text{ex}$ are easily described by
adding a term $-q_i^\text{br} V_i^\text{ex}$ to the Lagrangian, where
$q_i^\text{br}$ is the corresponding branch charge expressed in terms of the
loop charges.  Similarly, for a loop $l$ with loop charge $Q_l$ and external
flux $\Phi^\text{ex}_l$ which generates a positive voltage drop $ V = \dot
\Phi^\text{ex}_l$ in the loop current direction, a term $\dot{Q}_l
\Phi^\text{ex}_l$ should be added to the Lagrangian.

Offset charges are slightly more difficult to handle since they modify the
current balance rather than the voltage balance.  This means that they cannot
be described in terms of loop charges with the simple rules given in
Sec.~\ref{sec:circuit_quantization_loop_charges} since no term added to the
equations of motion can modify the current balance.  Instead, one must
represent them through additional branches which are described in terms of
node fluxes.  This requires a mixed loop charge/node flux formulation that we
will describe in detail in Sec.~\ref{sec:mixed_formulation}.  In the end,
however, we obtain a simple rule that we will state now for convenience and
whose proof we defer to Sec.~\ref{sec:mixed_formulation}.  To understand the
rule, we first note that the lumped element description requires overall
charge neutrality since otherwise there is a net electric field that extends
through the circuit and is not confined to the lumped elements.  This means
that we can only specify $n-1$ offset charges $q^\text{ex}_i$ with $i\neq 0$
on the $n$ islands of the circuit since overall neutrality implies that the
offset charges leave behind a charge $q_0^\text{ex}= -\sum_{i\neq 0}
q_i^\text{ex}$ on the ground node with $i=0$.

\begin{figure}[tb]
\centering
\includegraphics{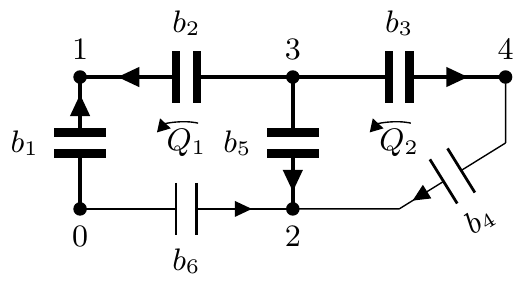}
\caption{Example network consisting of 6 branches $b_1,\dots,b_6$ and $5$
  nodes $0,\dots,4$. If the branches $b_4$ and $b_6$ (thin lines) are removed
  from the graph, the branches $b_1, b_2, b_3, b_5$ (thick lines) still
  connect all nodes and therefore form a spanning tree of the graph. Choosing
  the node $0$ as the ground node, we can only specify the offset charges
  $q_1^\text{ex}, \dots, q_4^\text{ex}$ on the remaining nodes since the
  ground node must carry the charge $q_0^\text{ex}=-\sum_{i=1}^4
  q_i^\text{ex}$ to guarantee overall charge neutrality. As explained in the
  main text, in order to accommodate the offset charges in our circuit
  description, we have to determine which offset charges are transported
  through which tree branches on their way from the ground to their respective
  node.  For example, the offset charge $q_2^\text{ex}$ has to be transported
  along the tree branches $b_1$, $b_2$ and $b_5$ in order to arrive at node
  $2$.\label{fig:offset_charges_tree}}
\end{figure}

In order to handle the offset charges $q_i^\text{ex}$, one must consistently
keep track of the paths through which the polarization charge propagates on
its way from the ground node to node $i$.  To that end, we use the concept of
a spanning tree.  For a graph, a spanning tree is defined as a subgraph which
does not have any loops and connects all nodes.  The branches of the graph
that belong to the spanning tree are called tree branches.  Since a spanning
tree of a connected graph with $n$ nodes has $n-1$ tree branches, we obtain a
one-to-one relation between the $n-1$ tree branches and the $n-1$ offset
charges. 

The offset charges can now be included following a number of simple steps.  We
first choose a ground node and construct a spanning tree of the circuit.  In a
second step, we express the branch charges $\vec q^\text{br}$ of the circuit
as differences of loop charges, following the same reasoning that we apply in
absence of offset charges.  As a last step, for all tree branches $b$, we
shift the resulting charge expression $q^\text{br}_b$ by replacing
$q^\text{br}_b \mapsto q^\text{br}_b \pm \Sigma^\text{ex}_b$.  We use the plus
sign if the branch $b$ is directed away from the ground node and the minus
sign otherwise.  The sum $\Sigma^\text{ex}_b$ is the sum of all the external
offset charges $q^\text{ex}_i$ that have passed through the tree branch $b$ on
their unique way from the ground node to node $i$ (within the tree).  Note
that the specific choice of spanning tree is a gauge in the sense that it has
no physical consequences.  It only amounts to a redefinition of the meaning of
the charges $\vec{q}^\text{br}_b$ that no longer give the physical charge on
the respective tree element.

As an example, consider the capacitive network depicted in
Fig.~\ref{fig:offset_charges_tree} consisting of six branches $b_1,\dots,b_6$
and 5 nodes $0,\dots,4$ with respective offset charges
$q_1^\text{ex},\dots,q_4^\text{ex}$.  As a
first step, we choose the node $0$ as the ground node and use a spanning tree
consisting of the branches $b_1$, $b_2$, $b_3$, and $b_5$ (thick lines).  For
the next steps, let us explicitly consider the branch $b_1$.  In absence of
offset charges, the branch charge $q_1^\text{br}$ can be expressed as
$q_1^\text{br} = - Q_1$ in terms of loop charges.  Next we determine
$\Sigma_1^\text{ex}$.  Since the offset charges $q_1^\text{ex}$,
$q_2^\text{ex}$, $q_3^\text{ex}$, and $q_4^\text{ex}$ all have to pass through
the branch $b_1$ in order to reach their respective nodes while traversing
only tree branches, we find $\Sigma_1^\text{ex} = \sum_{i=1}^4 q_i^\text{ex}$.
Since $b_1$ is directed away from the ground node, including the offset
charges amounts to the replacement $q_1^\text{br} = -Q_1 \mapsto -Q_1 +
\Sigma_1^\text{ex}$.  Proceeding in a similar way with the other branches, we
obtain the Lagrangian
\begin{align}
\mathcal L&=
 {-}\frac{(q_1^\text{ex}{+}q_2^\text{ex}{+}q_3^\text{ex}{+}q_4^\text{ex}{-}Q_1)^2}{2 C_1}
 - \frac{(Q_1 {-}q_2^\text{ex}{-}q_3^\text{ex}{-}q_4^\text{ex})^2}{2 C_2}
 \nonumber \\
&  
 - \frac{(q_4^\text{ex}{-}Q_2)^2}{2 C_3} - \frac{Q_2^2}{2 C_4}
-\frac{(q_2^\text{ex}{+}Q_2{-}Q_1)^2}{2 C_5} - \frac{Q_1^2}{2 C_6}. 
\end{align}
We note that in line with our previous discussion, the charge expressions of the
branches $b_4$ and $b_6$ which do not belong to the tree have not been modified
by the offset charges. 

With the offset charge description, we can simply represent a current source,
which injects a current $I^\text{ex}$ into the circuit and points from node
$n$ to node $n'$ by adding the offset charge $\int^t d t' I^\text{ex}(t')$ at
node $n'$ and the offset charge $-\int^t d t' I^\text{ex}(t')$ at node $n$.

\section{Duality between node fluxes and loop charges}

\label{sec:duality}

In the previous section, we have discussed two representations of the
Lagrangian of a circuit, one in terms of node fluxes and the other in terms of
loop charges.  In the following, we will call such a change in description  of
the \textit{same} system from node fluxes to loop charges a \textit{passive}
duality transformation.  Besides those passive duality transformations of the
same circuit, one can also consider \textit{active} duality transformations
which yield a different, electromagnetically dual circuit whose charge dynamics
is identical to the flux dynamics of the original circuit or vice-versa.
Electromagnetic circuit dualities have been discussed on a per-case basis in
the mesoscopic physics literature~\cite{mooij:06, guichard:10, kerman:13} but,
to our knowledge, a general construction scheme has not been spelled out so
far. 

In this section, we will explain how to explicitly construct both passive and
active duality transformations with the help of loop charges.  We will start by
discussing the explicit construction of passive duality transformations.
Previously, we have focused on the question on how to read off the appropriate
Lagrangian in either representation directly from a given circuit graph.   We
now show how one can transform one representation into the other.  While this
is of technical interest, we want to highlight that this subsection may be
skipped on first reading since in practice it is sufficient and much easier to
use the rules given in Sec.~\ref{sec:circuit_quantization_loop_charges} for the
construction of the loop charge Lagrangian.  We proceed by outlining in
Sec.~\ref{sec:duality_active} a straightforward way of constructing
electromagnetic circuit dualities using loop charges.

\subsection{Passive duality transformations}

\label{sec:duality_passive}

The transformation from the node flux to a loop charge representation is
particularly easy to perform in the path integral picture~\cite{hibbs}.  In
this case, the unitary time-evolution operator $e^{-i H t/\hbar}$ is
represented in the form
\begin{align}\label{eq:time_evolution_pathint}
e^{-i H t/\hbar} \rightarrow \int\! \mathcal{D}[\vec \phi(t)]\, 
e^{(i/\hbar) \int^t\!dt'\, \mathcal{L}(\vec \phi^\text{br})}, 
\end{align}
where the path-integration is performed over the $n-1$ node fluxes of the
circuit graph with $n$ nodes.  Note that we have also suppressed the
dependence of the Lagrangian on $\dotvec{\phi}^\text{br}$ for brevity.  The
description in terms of branch fluxes $\vec \phi^\text{br}$ is linked to a
description in terms of branch charges $q_b^\text{br} = \partial
\mathcal{L}(\vec \phi^\text{br}, \dotvec{ \phi}^\text{br})/\partial \dot
\phi_b^\text{br}$ through the Legendre transformation.  For the following, it
will be convenient to perform this Legendre transformation in a slightly more
general form through the Fourier transformation
\begin{align}\label{eq:legendre_fourier_transform}
  e^{(i/\hbar) \int^t\!dt'\, \mathcal{L}(\vec \phi^\text{br})} 
= \int \!\mathcal{D}[\vec q^\text{br}(t)] 
e^{(i/\hbar) \int^t\!dt' [ 
	\tilde{\mathcal{L}}(\vec{q}^\text{br})
	- \vec{q}^\text{br}\cdot \dotvec{\phi}^\text{br}]}, 
\end{align}
where the Lagrangian $\tilde{\mathcal{L}}(\vec q^\text{br}, \dotvec
q^\text{br})$ is defined implicitly such that
Eq.~\eqref{eq:legendre_fourier_transform} holds.  At the saddle-point level or
for a Lagrangian $\tilde{\mathcal{L}}(\vec q^\text{br}, \dotvec{q}^\text{br})$
that is quadratic in its arguments, performing the $\vec q^\text{br}$
integration shows that $\mathcal{L}(\vec \phi^\text{br},
\dotvec{\phi}^\text{br})$ is simply the Legendre transformation of $\tilde
{\mathcal L}(\vec q^\text{br}, \dotvec{q}^\text{br})$.

To proceed further, we need to relate the node fluxes $\vec \phi$ to the
branch fluxes $\vec \phi^\text{br}$.  For this, we make use of the basis
node-edge incidence matrix $A$ which is a $\mathbb{R}^{(n-1) \times b}$
matrix for the $n-1$ nodes fluxes and the $b$ branches.  Its entries $A_{ij}
\in \{1,-1\}$ indicate whether the branch $j$ enters ($-1$) or leaves $(+1)$
node $i$.  It allows us to express the Kirchhoff current law in the form $A
\dotvec{q}^\text{br} = 0$ and it relates the branch and node fluxes via $\vec
\phi^\text{br} = A^T \vec \phi$.

Performing a partial integration on the term $-\vec q^\text{br} \cdot
\dotvec{\phi}^\text{br} = -\vec q^\text{br} \cdot A^T \dotvec{\phi}$ in the
exponent of expression~\eqref{eq:legendre_fourier_transform}, inserting the
resulting expression into Eq.~\eqref{eq:time_evolution_pathint}, and performing
the integration over $\vec \phi$ results in a constraint:
\begin{align}
e^{-i H t/\hbar} \rightarrow \int\! \mathcal{D}[\vec q^\text{br}(t)] 
e^{(i/\hbar) \int^t\!dt'\,
  \tilde{\mathcal{L}}(\vec{q}^\text{br})} \delta[A \dotvec{q}^\text{br} (t)],
\end{align}
where the $\delta$ function has to be understood in such a way that it
demands the vanishing of its argument at each point in time.  The constraint
$A \dotvec{q}^\text{br} = 0$ is of course nothing but the Kirchhoff current
law.  As we have discussed in details in
Sec.~\ref{sec:circuit_quantization_loop_charges}, we can guarantee the
Kirchhoff current law for a planar circuit by considering loop charges.  This
resolves the constraint and we obtain the dual representation
\begin{align}\label{eq:dual_path_integral}
e^{-i H t/\hbar} \rightarrow  \int\! 
\mathcal{D}[\vec Q(t)] e^{(i/\hbar) \int^t  
\!dt'\,	\tilde{\mathcal{L}}[\vec{q}^\text{br}(\vec Q)]} 
\end{align}
in terms of loop charges. For the convenience of the reader, we repeat this
derivation in a slightly more rigorous way in App.~\ref{app:duality}.

We have thus explicitly constructed the passive duality transformation linking
a representation in terms of node fluxes to a representation in terms of loop
charges. We want to stress once again that in practice it is
much easier and much less error-prone to perform the construction of the
circuit Lagrangian using the rules explained in details in
Sec.~\ref{sec:circuit_quantization_loop_charges}, rather than starting with a
node-flux representation and repeating the calculation outlined above.

It is interesting to note that the duality transformation used here is
essentially the same as the one used in the analysis of the classical
two-dimensional XY-model~\cite{savit:80} or the Schmid-Bulgadaev transition.
In fact, the analogy to the XY-model suggests that Josephson junctions can be
described in the loop charge formulation by making the Villain approximation
for the cosine dispersion $E_J \cos(2\pi \phi^\text{br}/\Phi_Q)$ of a
Josephson junction with branch flux $\phi^\text{br}$.  There, one replaces the
cosine dispersion by the function $-\min_{m \in \mathbb{Z}} E_J ( 2\pi
\phi^\text{br}/\Phi_Q - 2\pi m)^2/2$ which retains the $\Phi_Q$ periodicity
while being quadratic in $\phi^\text{br}$.  This allows to perform the path
integration over $\vec \phi$ and construct a charge-based description of a
Josephson junction in the Villain approximation.  We will not pursue this idea
further since we will introduce in Sec.~\ref{sec:applications} an alternative
way to describe a Josephson junction (using loop charges) that is based on the
adiabatic separation of the (fast) Cooper-pair transport through the junction
and the (slow) transport of polarization charge through the rest of the
circuit.

\subsection{Active duality transformations: electromagnetic circuit duality}
\label{sec:duality_active} 
In the previous section, we have explained the representations of circuits in
terms of node fluxes or loop charges which are related by a passive duality
transformation.  We now want to show that loop charges are also useful for
constructing active duality transformations.  Specifically, given a graph $g$
of a circuit that is described in terms of node fluxes and has a corresponding
Lagrangian $\mathcal{L}(\vec \phi, \dotvec{\phi})$, we define its
electromagnetically dual circuit with graph $G$ as the circuit whose
description in terms of loop charges yields a Lagrangian that is of the same
form as $\mathcal{L}(\vec \phi, \dot{\vec \phi})$ with $\vec \phi$ replaced by
a vector of loop charges $\vec Q$.  We will see below that a dual circuit
exists for planar circuits which are effectively two-dimensional such that the
closure of flux lines in the third dimension can be ignored; this is in
contrast to classical electromagnetism where electromagnetic dualities only
exist in vacuum due to the absence of magnetic monopoles~\cite{jackson}.

\begin{figure}[tb]
\centering
\includegraphics{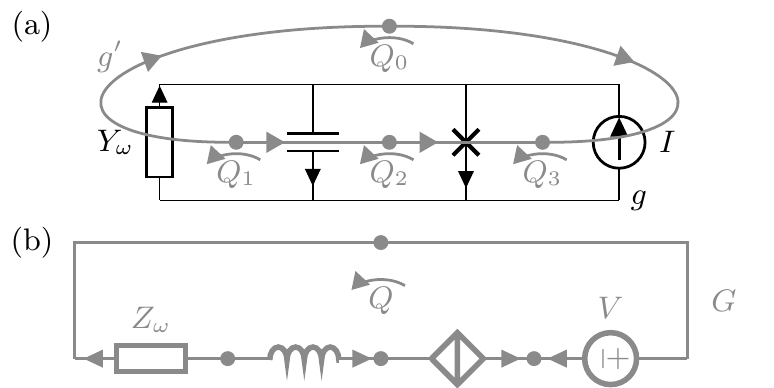}
\caption{In (a), we illustrate the construction of the graph $g'$ dual to a
	graph $g$. As explained in the main text, we construct $g'$ by placing
	one node into each loop of $g$. We connect two nodes in the dual graph
	$g'$ whenever there is a circuit element on branch $b$ in $g$ that
	separates the
  corresponding loops $l_0$ and $l_1$. The orientation on the branch in $g'$
  is chosen such that the branch points towards $l_1$ if the orientation of
  the loop charge $Q_{l_1}$ is consistent with the orientation of the original
  branch in $g$ and away from $l_1$ otherwise. In (b), we show the
  electromagnetic dual graph $G$ that is obtained from $g'$ by replacing the
	elements according to the rules given in
	Fig.~\ref{fig:electromagnetic_duality_conversion}.
	\label{fig:electromagnetic_dual}}
\end{figure}

In order to construct the dual circuit $G$, we first need the notion of a dual
graph $g'$.  In the node flux formulation, each branch flux $\phi^\text{br}_b$
is obtained as the difference of precisely two node fluxes.  We have seen
previously that for a planar circuit described in terms of loop charges, we can
similarly describe each branch charge $q^\text{br}_b$ as the difference of two
loop charges, provided we also place a loop charge $Q_0 = 0$ at the exterior of
the circuit.  For this reason, we construct the dual graph $g'$ by placing one
node into each loop of the original graph $g$, including the ``loop'' at the
exterior~\cite{Note2}. For each branch $b$ representing a circuit element that
is common to the loops $l_0$ and $l_1$, we add a branch $b'$ in the dual graph
representing the same circuit element that joins the dual nodes at $l_0$ and
$l_1$.  We choose the orientation of the dual branch such that it points
towards $l_1$ if the orientation of the original branch is consistent with the
loop charge orientation $Q_{l_1}$ and away from $l_1$ otherwise.  This gives a
consistent scheme provided we choose a counter-clockwise orientation for all
loop charges as described in Sec.~\ref{sec:circuit_quantization_loop_charges}.
The construction scheme is illustrated in
Fig.~\ref{fig:electromagnetic_dual}(a) for a simple circuit.  Associating the
loop charges of the original circuit with the nodes of the dual graph, the
charge on branch $b$ of the original circuit can be obtained  as the negative
(discrete) gradient of the loop charges along the branch $b'$ of the dual
graph.  Up to a sign, we thus obtain the branch charge $q_b$ of the original
graph $g$ from the dual graph $g'$ in a way that is completely analogous to the
node flux formulation.  Note that the dual graph does not represent a lumped
element representation of a physical circuit but it should rather be considered
a handy mnemonic for the loop charge representation of the original circuit.
We highlight that iterating this procedure twice gives back the original graph
with the orientation of all branches reversed.

To construct the dual circuit $G$, we start by considering the dual graph $g'$
of $g$ as a lumped element representation of an actual circuit different from
the original circuit.  As we have explained before, we can understand the loop
charge formulation of $g'$ by thinking about the loop charges of $g'$ sitting
on the nodes of $(g')'$.  Now, since $(g')'$ is just the original graph $g$
with all branch orientations reversed, we effectively obtain the branch
charges of $g'$ as the gradient of the loop charges positioned on the nodes of
the original graph $g$.  Thus, we obtain the result that the node fluxes of
$g$ are in one to one relation with the loop charges of $g'$.  From $g'$, we
obtain the dual circuit $G$ by replacing circuit elements of $g'$ in such a
way that the loop charges of $G$ have the same dynamics as the node fluxes of
$g$.  In order to have the same dynamics, the terms in the Lagrangian
corresponding to the circuit elements have to be equal (up to interchanging
$\bm \varphi$ with $\bm Q$).  For example, a capacitive element in $g$
corresponds to a (kinetic) term of the form $C \dot{\phi}^2/2$ and its dual is
thus given by an inductor $L\dot Q^2/2$ (which leads to a kinetic term in the
loop charge description).  More generally, we obtain the electromagnetically
dual circuit $G$ from the dual graph $g'$ of $g$ by replacing all elements in
$g'$ according to the rules given in
Table~\ref{fig:electromagnetic_duality_conversion}.  This procedure is
illustrated in Fig.~\ref{fig:electromagnetic_dual}(b) for a simple circuit.

\begin{table}[tb]
\centering
\begin{tabular}{cp{1em}c}
  \hline
  Original & & Dual \\
  \hline
  Capacitance $C$ & &Inductance $L$ \\
  Josephson junction $E_J$ & & Phase-slip junction $E_S$ \\
  Flux $\phi^\text{ex}$ through loop $n$ & &Offset charge $q^\text{ex}$ at node
  $n$ \\
  Voltage source $V$ & & Current source $I$\\
  Admittance $Y_\omega$ & &Impedance $Z_\omega$ \\\hline\hline
\end{tabular}
\caption{Circuit elements and their corresponding elements in the
electromagnetically dual circuit. 
\label{fig:electromagnetic_duality_conversion}}
\end{table}
\section{Dissipation and environments}\label{sec:dissipation}

So far, we have analyzed \emph{closed systems} where the energy is conserved.
We have given a recipe to calculate the Lagrangian $\mathcal{L}(\vec Q,
\dotvec Q)$ that corresponds to a specific lumped-element circuit.  In a
typical application, we would then go on by introducing the Hamiltonian and
canonically quantizing position $\vec Q$ and momentum $\vec \Phi =
\partial_{\dot{\vec Q}} \mathcal{L}$.  Given an initial configuration
$\Psi_{t_0}(\vec Q)$, we obtain a wavefunction $\Psi_t(\vec Q)$ that
describes the evolution of the probabilities $|\Psi_t(\vec Q)|^2$ to find the
system in a specific state $\vec Q$ at time $t$.

An altogether different but equivalent approach is the path integral
method~\cite{hibbs}, which we already briefly discussed in
Sec.~\ref{sec:duality_passive}.  There the wavefunction is
obtained by the expression
\begin{equation}\label{eq:path-integral}
  \Psi_t(\vec Q) = \int\! \mathcal{D}[\vec Q(t)] 
  e^{(i/\hbar) \int_{t_0}^t \!dt'\, \mathcal{L}} \,\Psi_{t_0}(\vec Q')
\end{equation}
that sums over all paths $\vec Q(t)$ fulfilling the boundary conditions $\vec
Q(t_0) =\vec Q'$ and $\vec Q(t) =\vec Q$.  Note that in this approach there is
neither a need to go over to a Hamiltonian nor  to postulate
canonical quantization rules.

In conventional electronics, there are elements called resistors that do not
conserve energy.  In a quantum setting, this corresponds to \emph{open
systems}, i.e., a system coupled to an environment; an example is an
electronic circuit which is coupled to the outside via a electromagnetic
transmission line.  We note that recently there has been a lot of progress in
quantizing general linear environments in terms of a few relevant degrees of
freedom~\cite{nigg:12, bourassa:12, leib:12, solgun:14, solgun:15}. Here, we
will describe the environment as an effective action on the system degrees of
freedom. 

In the theory of open systems, the interest is in characterizing the system
in questions without having to specify the full wavefunction of the system
together with its environment.  As in this case the system does not stay
in a pure state, it necessarily has to be characterized by its density matrix
$\rho_t(\vec Q^+, \vec Q^-)$ whose diagonal elements give the probability to
observe the system in a particular state $\vec Q^- =\vec Q^+$ and the
off-diagonal terms characterize the coherences.  We see that the fact that the
system is open requires to double the degrees of freedom, i.e., going from
$\vec Q$ to $\vec Q^\pm$. The dynamics of the system is simply given by
\begin{equation}\label{eq:keldysh}
  \rho_t(\vec Q^+, \vec Q^-) =  \int \!
  \mathcal{D}[\vec Q^{+}(t), \bm Q^{-}(t)] 
  e^{i \mathcal{S}/\hbar} \,\rho_{t_0}(\vec Q'^+ ,\vec Q'^-)
\end{equation}
where $\mathcal{S} = \mathcal{S}_S + \mathcal{S}_E$ has a contribution due to
the system (without the environment)
\begin{equation}\label{eq:S_S}
  \mathcal{S}_S = \int_{t_0}^t\! dt' 
  [\mathcal{L}(\vec Q^+, \dot{\vec Q}^+,t) -
  \mathcal{L}(\vec Q^-, \dot{\vec Q}^-,t) ].
\end{equation}
The influence of the environment can be captured by the so-called influence
functional $\mathcal{S}_E$~\cite{feynman:63}.

If the environment is a linear system in equilibrium characterized by the
impedance $Z_\omega$, the influence functional can be calculated
explicitly~\cite{caldeira:83,schon:90}.  If the branch $b$ (between the two
loops $l_0$ and $l_1$) with branch charge $q^\text{br}_b = Q_{l_1} - Q_{l_0}$
is shunted by the impedance $Z_\omega$, we obtain the additional action
$\mathcal{S}_E = \mathcal{S}_R + \mathcal{S}_D$
with a reactive part
\begin{equation}\label{eq:S_R} 
  \mathcal{S}_R = \int\!\frac{d\omega}{4\pi}
  \mathop{\rm Im}(Z_\omega) \, \omega \Bigl( | \tilde Q^-_{\omega}|^2 -  |
  \tilde Q^+_{\omega}|^2 \Bigr)
\end{equation}
where the Fourier-transform  $\tilde Q^\pm_{\omega} = \int_{t_0}^t\! dt'\,
q^{\text{br},\pm}_{b}(t') e^{i\omega t'}$ enters.  Note that in the reactive
part, similar to the system, the variables $\tilde Q^+$ and $\tilde Q^-$ are
not coupled which corresponds to the fact that the evolution of the ket and
bra in a pure state $\rho_{t} = \Psi_t(\vec Q^+) \Psi^*_t(\vec Q^-)$ are
independent of each other.  In particular, for a simple inductance $L$ with
impedance $Z_\omega = -i \omega L$ or a capacitance $C$ with impedance
$Z_\omega = i/\omega C$, the expression~\eqref{eq:S_R} reproduces the results
of Fig.~\ref{fig:loop_charge_summary}.

The dissipation destroys this factorization and makes the doubling of the
degrees of freedom inevitable. In fact, it is useful to introduce new
variables $\tilde{Q}^\text{cl}_{\omega} = \tfrac12 (\tilde Q^+_{\omega}
+\tilde Q^-_{\omega}  )$ and $\tilde Q^q_{\omega} = \tilde Q^+_{\omega}
-\tilde Q^-_{\omega}  $ in terms of which the dissipative part of the  action
reads
\begin{equation}\label{eq:S_D}
  \mathcal{S}_D =  \int\!\frac{d\omega}{2\pi}
  \mathop{\rm Re}(Z_\omega) \, \omega \Bigl[
    \mathop{\rm Im}(\tilde Q^\text{cl}_{-\omega} \tilde Q^q_\omega) +
  i (2 n_\omega + 1)  
  | \tilde Q^q_{\omega}|^2 \Bigr];
\end{equation}
here, $n_\omega$ denotes the occupation probability of the  mode at frequency
$\omega$ in the environment. In particular, in equilibrium, we have the
Bose-Einstein distribution $n_\omega = {(e^{\hbar \omega/k_B T} -1)}^{-1}$.
The two terms in~\eqref{eq:S_D} have different tasks: the first term
introduces dissipation in the equation of motion and the last term leads to
fluctuations, see also below.

As an example, we would like to analyze a setup where a phase-slip junction in
series with an inductor and a resistance is voltage biased at voltage $V_0$,
which is illustrated in Fig.~\ref{fig:electromagnetic_dual}(b).  The circuit
consists of a single loop with loop charge $Q$.  This system is the dual of
the resistively-shunted Josephson junction shown in
Fig.~\ref{fig:electromagnetic_dual}(a).\cite{tinkham} The Lagrangian assumes
the form
\begin{equation}\label{eq:lag_shunt}
  \mathcal{L} =  \frac{L \dot Q^2}{2}  + E_S \cos( \pi Q/e) + V_0 Q
\end{equation}
involving both the phase-slip junction as well as the voltage bias. The action
of the system is obtained via~\eqref{eq:S_S}. The Ohmic resistance  is
modelled by dissipative action~\eqref{eq:S_D} with $\mathop{\rm Re}(Z_\omega)
=R$. 

How the system dynamics is modified by dissipation depends on temperature.
Let us first consider the case $T = 0$, which can be analyzed using the
well-known results for the dual problem of the resistively-shunted Josephson
junction.  For the following, we consider the case $V_0 = 0$.  It is then
advantageous to decompose the total flux within the loop in the form
$\phi+\Phi$ with $\phi \in [0, \Phi_Q]$ and $\Phi/\Phi_Q \in \mathbb{Z}$.  The
former flux can be interpreted as the Bloch momentum associated with the
dynamics of $Q$ in the $2e$-periodic potential due to the phase-slip junction,
while the latter is connected to the dynamics within a single unit cell of
size $2e$.  For zero shunt resistance, $R = 0$, the flux $\phi$ (Bloch
momentum) is conserved, corresponding to a complete delocalization of $Q$ over
the valleys of the cosine potential.  Localizing the charge $Q$ in a single
valley of the periodic potential requires a superposition of all Bloch momenta
$\phi$.  The fluctuation-dissipation theorem, $S_\phi(\omega) \propto \Re
(Z_\omega)$, shows that increasing $\Re (Z_\omega)$ will increase the
fluctuations of $\phi$ at frequency $\omega$ as described by the spectral
density $S_{\phi}(\omega) = \int \!dt \, e^{i \omega t} \langle \phi(t)
\phi(0)\rangle$. This suggests that for $R$ sufficiently large such that the
fluctuations of $\phi$ exceed $\Phi_Q$, $Q$ will eventually localize within a
single valley of the periodic potential.  The transition from a state
delocalized over different valleys of the periodic potential to a localized
state is known as the Schmid-Bulgadaev quantum phase transition that was
mainly studied in the dual problem of the resistively shunted Josephson
junction (for zero current
bias)~\cite{schmid:83,bulgadaev:84,guinea:85,schon:90}. Translated to our
problem, the results imply that $Q$ is localized for $R > R_Q$ and remains
delocalized for $R < R_Q$.

For finite temperature $T$, the Schmid-Bulgadaev transition is formally absent
because thermal activation will always lead to a finite probability for the
charge $Q$ to transition between different valleys of the
potential~\cite{schon:90}. However, as long as we are on the insulating side
of the Schmid transition with $R > R_Q$ where quantum tunneling of $Q$ is
absent, we can describe the dynamics of $Q$ semi-classically.  This
corresponds to expanding the action around $Q^q = 0$~\cite{kamenev}, which
leads to
\begin{multline}\label{eq:S}
  \mathcal{S} = \int_{t_0}^t\! dt' [V_0 - L \ddot{Q}^\text{cl} - R 
    \dot{Q}^\text{cl} - V_c \sin(\pi
  Q^\text{cl}/e) ] Q^q \\
  +iR \int\! \frac{d\omega}{2\pi} \omega  (2n_\omega + 1)
  |Q_\omega^q|^2
\end{multline}
with $V_c = \pi E_S/e$. Next, we introduce the fluctuation $\xi$ of the
voltage over the resistor via a Hubbard-Stratonovich transformation. In fact,
we have that
\begin{equation}\label{eq:hubbard}
  e^{i \mathcal{S}_D} = \int\! \mathcal{D}[\xi(t)] \exp\biggl[- \!\int\!
  \frac{d\omega}{2\pi} \Bigl(i
  \xi^*_{\omega} Q^q_\omega  + \frac{|\xi_\omega|^2}{4R
  \omega (2n_\omega+1)}  \Bigr)\biggr].
\end{equation}
After this transformation, the action is linear in $Q^q$ which allows for
performing the path-integral over $Q^q$. The result is the Langevin equation
\begin{equation}\label{eq:phaseslipjunction_langevin}
   V_0 - L \ddot{Q}^\text{cl} - R 
    \dot{Q}^\text{cl} - V_c \sin(\pi
  Q^\text{cl}/e)  = \xi(t)
\end{equation}
for $Q^\text{cl}(t)$.  In the end, as $Q^q= Q^+ - Q^-$ is small, we obtain a
result for the time-evolution of the probability distribution $P_t(Q) = \rho_t
(Q,Q)$; with $Q=Q^\text{cl} \approx Q^+ \approx Q^-$. It is given by
\begin{equation}\label{eq:prob}
  P_t (Q) = \int\! \mathcal{D}[\xi(t)]   \exp\biggl[ - \! \int \!
    \frac{d\omega\,|\xi_\omega|^2}{8\pi R
  \omega (2n_\omega+1)} \biggr] P_{t_0}(Q')
\end{equation}
where  $Q^\text{cl}(t)$ fulfills the Langevin equation with $Q^\text{cl}(t_0)=
Q'$ and $Q^\text{cl}(t)= Q$. In particular, the fluctuating part of the
voltage $\xi(t)$ is Gaussian with mean $\langle \xi \rangle =0$ and variance
\begin{equation}\label{eq:fluct}
  \langle \xi_{\omega'} \xi_\omega \rangle
= 4 \pi R \omega \coth(\hbar \omega/ 2k_B T) \delta(\omega' + \omega)
\end{equation}
where we used the fact that $2 n_\omega+ 1 = \coth(\hbar \omega/2 k_B T)$ in
equilibrium.

\section{Mixed circuit quantization and proof of circuit rules}\label{sec:mixed_formulation} 
In the previous section, we have reviewed the node flux description and
explained in some detail the loop charge description of circuits.  We now want
to show that one can also combine both descriptions such that part of the
circuit is described in terms of node fluxes while the other is described in
terms of loop charges.  As an example,  we  will use this approach to prove
the rules for the inclusion of offset charges given above.

Let us assume that we decide to describe a only a certain subset of the
branches of the graph in terms of loop charges.  In the following, we will
refer to the part of the graph spanned by the corresponding branches as the
subgraph, while the remaining branches belong to what we will call the
subgraph complement.  The boundary nodes of the subgraph are the nodes that
possess both incident branches that belong to the subgraph as well as incident
branches that belong to its complement.  We denote the vector of node fluxes
at the boundary nodes by $\vec \phi^\partial$.  Similarly, the boundary loops
of the subgraph with loop charges denoted by $\vec Q^\partial$ are the loops
with branches that partly belong to the subgraph and partly belong to its
complement.  Since the voltage drops over the branches to which the boundary
loops belong as well as the currents in the branches incident on the boundary
nodes are partly described in terms of node fluxes and partly in terms of loop
charges, the Kirchhoff voltage law at the boundary loops and the Kirchhoff
current law at the boundary nodes is no longer automatically fulfilled.  We
therefore have to ensure it manually by adding appropriate terms to the
Lagrangian.  Let us denote the current flowing from a boundary node $i$ to a
neighboring node $j$ within the subgraph by $\dot q_{ij}$.  Since the
Euler-Lagrange equations with respect to the node flux $\phi_i$ yield the
currents flowing away from node $i$, we can ensure the Kirchhoff current law
by adding the term $-\sum_i \phi^\partial_i \sum_j \dot q_{ij}$ to the
Lagrangian.  Similarly, for the boundary loops with charges $Q^\partial_i$, we
can guarantee the Kirchhoff voltage law by adding a term $-\sum_i Q^\partial_i
\sum_j \dot \phi_{ij}$, where $\dot\phi_{ij}$ are the voltage drops (in the
loop current direction) over the parts of the loop that are in the subgraph
complement.

The first of the terms just described manifestly guarantees current
conservation while the second manifestly guarantees the Kirchhoff voltage law.
Importantly, both terms are identical up to a total time derivative, as we
show in App.~\ref{app:mixed_formulation_current_conservation}.  As a
consequence, if one wants to guarantee both the Kirchhoff current law as well
as the Kirchhoff voltage law, we have to add one (and only one) of them to the
circuit Lagrangian.

\begin{figure}[tb]
\centering
\includegraphics{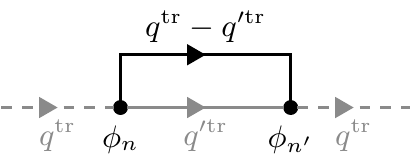}
\caption{In order to describe the presence of offset charges, a virtual branch
	(solid black line) representing the effect of the displacement
	currents is added in parallel to each tree branch of the original
	circuit (solid gray line).  As a consequence, the charge $q^\text{tr}$
	entering the tree branch splits into the charge $q^{\prime
		\text{tr}}$ on the tree element and the charge $q^\text{tr} -
	q^{\prime \text{tr}}$ on the virtual branch.  The current flowing away
	from node $n$ and $n'$ into the subgraph (gray) is given by
	$\pm(q^{\prime \text{tr}} - q^\text{tr})$.  Ensuring the Kirchhoff
	laws therefore requires adding the terms $-(\phi_n - \phi_{n'})(\dot
	q^{\prime \text{tr}} - \dot q^\text{tr}) = -\phi^\text{tr} ( \dot
	q^{\prime \text{tr}} - \dot q^\text{tr})$ with the tree branch flux
	$\phi^\text{tr} = \phi_{n} - \phi_{n'}$ to the
	Lagrangian.\label{fig:offset_charges}}
\end{figure}

Let us now use these results to prove the rules for the inclusion of offset
charges described in Sec.~\ref{sec:circuit_quantization_loop_charges}.  As we
have discussed there, offset charges must be modeled through the inclusion of
additional lumped elements in the circuit.  These elements are naturally
described in terms of node fluxes since they modify the current balance.
Therefore, in order to describe the presence of offset charges $\vec
q^\text{ex}$ on the nodes of the circuit, we add to each of the tree branches
with charges $\vec{q}^\text{tr}$ another virtual parallel branch which will
represent the action of the displacement currents and will be described in
terms of node fluxes.  As a consequence, only a fraction $\vec{q}^{\prime
	\text{tr}}$ of the total charge $\vec{q}^\text{tr}$ entering the
branches will remain on the original tree element, while the charge $\vec
q^\text{tr} - \vec q^{\prime \text{tr}}$ will reside on the virtual branch.
Since the equations of motion with respect to $\vec \phi$ yield the currents
flowing away from the respective nodes, offset charges $\vec q^\text{ex}$ on
the nodes of the circuit correspond to a term $\dotvec{\phi} \cdot \vec
q^\text{ex}$ in the Lagrangian.  In order to ensure the Kirchhoff laws, we
also have to add the terms $-\vec{\phi}^\text{tr} \cdot (\dotvec{q}^{\prime
\text{tr}} - \dotvec{q}^\text{tr}$), c.f.~Fig.~\ref{fig:offset_charges}.

We have already discussed in in Sec.~\ref{sec:duality_passive}
that the node-edge incidence matrix $A$ relates the branch fluxes and the node
fluxes as $\vec \phi^\text{br} = A^T \vec \phi$.  A decomposition of $\vec q^\text{br} =
(\vec q^\text{ch}, \vec q^\text{tr})$ into the vector of chord charges $\vec
q^\text{ch}$ and tree charges $\vec q^\text{br}$ gives rise to a corresponding
decomposition of $A = (A_\text{ch}, A_\text{tr})$ with $A_\text{tr}$ a square
matrix.  Since there are no loops in a tree, we have the result $A_\text{tr}
\vec v \neq 0$ for every vector $\vec v \in \mathbb{R}^b$, implying that
$A_\text{tr}$ has full rank and the inverse $A_\text{tr}^{-1}$ is
well-defined~\cite{*[{}][{, Theorem 2.2.}] chen}.  With the help of the matrix
$A$, we can write the expression added to the Lagrangian in the form
$\dotvec{\phi} \cdot \vec{q}^\text{ex} - \vec{\phi} \cdot A_\text{tr}
(\dotvec{q}^{\prime \text{tr}} - \dotvec{q}^{\text{tr}})$.  Since
$A_\text{tr}$ is invertible, the equations of motion with respect to $\vec
\phi$ yield the constraint $\dotvec{q}^{\prime \text{tr}} =
\dotvec{q}^{\text{tr}} - A_\text{tr}^{-1} \dotvec{ q}^\text{ex}$.  As a
result, we can simply ignore the virtual branches just introduced and continue
working with the original circuit graph, provided we simply replace each
expression in the Lagrangian involving the tree charge $\vec q^\text{tr}$ by
$\vec q^{\prime \text{tr}}$.  It can be shown that for all nodes $j$ that are
connected to ground through branch $i$, the entries of
$(A_\text{tr}^{-1})_{ij}$ are given by $\pm 1$ depending on whether branch $i$
points towards or away from ground, while they are zero for all other
nodes~\cite{chen}.  Using this, we reproduce the rules given previously.  We
show in the App.~\ref{app:mixed_formulation_current_conservation} that
proceeding similarly for a circuit with external fluxes that is described in
terms of node fluxes recovers the rules given in Ref.~\onlinecite{devoret:96}.

For completeness, we note that no such simple rule emerges if one intends a
mixed description of the circuit.  In that case, one does not get around
representing external fluxes and offset charges explicitly through virtual
additional circuit elements.  For an external flux $\Phi^\text{ex}_l$ in some
loop $l$ with loop charge $Q_l$ which is part of the subgraph or an offset
charge $q^\text{ex}_n$ at some node $n$ which is either part of the subgraph
complement or a boundary node, those virtual elements are easy to handle.  In
that case, they simply add the terms $\dot Q_l \Phi_l^\text{ex}$, $\dot \phi_n
q^\text{ex}_n$ to the Lagrangian without requiring additional terms to
guarantee the Kirchhoff laws.  For external fluxes in loops that lie
completely within the subgraph complement or offset charges at the nodes of
the subgraph (without the boundary nodes), however, the additional terms
guaranteeing the Kirchoff laws have to be added by hand.

\begin{figure}[tb]
\includegraphics[width=\columnwidth]{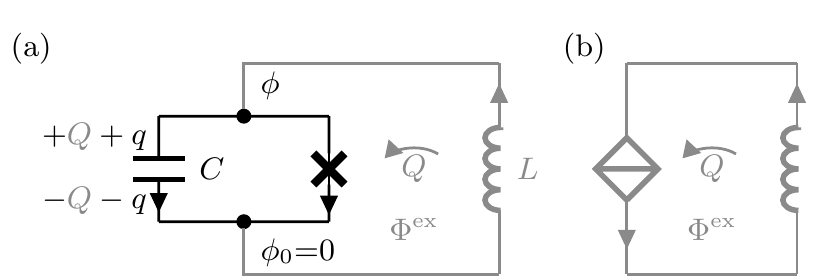}%
\caption{In (a), we show the idealized exact
fluxonium circuit and in (b), we show the approximate fluxonium representation
obtained after exploiting the passive duality transformation explained in
Sec.~\ref{sec:fluxonium}. The gray part of the circuit denotes the subgraph
described by loop charges.}%
\label{fig:fluxonium_circuit} %
\end{figure}

As an example, we consider the fluxonium circuit depicted in
Fig.~\ref{fig:fluxonium_circuit}.  We describe the inductive shunt in terms of
loop charges and the rest of the circuit in terms of node fluxes.  Using the
rules given above, we obtain the Lagrangian
\begin{align}\label{eq:fluxonium_lagrangian}
\mathcal{L} = \frac{C}{2} \dot \phi^2 + E_J \cos\biggl(\frac{2\pi \phi}{\Phi_Q}\biggr) +
\frac{L}{2} \dot Q^2 - Q \dot \phi + \dot{Q} \Phi^\text{ex}
\end{align}
with $E_J = \Phi_Q I_c/2\pi$.  Here, the first two terms are due to the
Josephson junction and its associated capacitance, which are described in
terms of node fluxes, while the term $L\dot Q^2/2$ is due to the inductive
shunt within the subgraph which is described in terms of loop charges.  The
voltage drop $\dot \phi$ in the direction of the loop charge $Q$ gives the
term $-Q \dot \phi$ guaranteeing the Kirchhoff voltage and current law.  The
external flux $\Phi^\text{ex}$ within the boundary loop adds the term $\dot Q
\Phi^\text{ex}$.  Since we describe the inductive shunt in terms of the
polarization charge $Q$, we have to take $\phi$ to be $\Phi_Q$-periodic since
only integer number of Cooper-pairs can flow from the ground to the node with
flux $\phi$ in absence of the inductive shunt.  Performing the Legendre
transformation, we obtain the Hamiltonian~\cite{apenko:89, zaikin:90}
\begin{align}\label{eq:fluxonium_ham_alternative}
  H_\text{flux} 
  = \frac{(q+Q)^2}{2C}  - E_J \cos\biggl(\frac{2\pi \phi}{\Phi_Q}\biggr) +
  \frac{(\Phi-\Phi^\text{ex})^2}{2 L},
\end{align}
where $q = \partial \mathcal{L}/\partial \dot \phi$ and the total flux $\Phi =
\partial \mathcal{L}/\partial \dot Q$ are canonically conjugate to $\phi$ and
$Q$. The Hamiltonian acts on wavefunction of the form $\psi(\phi,\Phi)$, where
$\phi$ is periodic (with period $\Phi_Q$).

The capacitive term of the Hamiltonian \eqref{eq:fluxonium_ham_alternative}
reveals that the physical charge $\tilde q = q + Q$ on the capacitor plate is
the sum of the charge $q \in 2e \mathbb{Z}$ (flowing through the Josephson
junction) and $Q$ (flowing through the inductor).  As the charges do not enter
individually, the operator $e^{i(\phi - \Phi)/\Phi_Q}$ commutes with the
Hamiltonian.  As a result, we obtain that the fluxes are equal with $\tilde
\phi = \Phi =\phi$, where the last equality holds modulo
$\Phi_Q$.\cite{Note3} We introduce the new
wavefunction
\begin{equation}\label{eq:bloch_theorem} 
  \tilde \psi(\tilde \phi)  = \psi(\tilde \phi, \tilde \phi)
\end{equation}
with $- i\hbar  \partial_{\tilde \phi} \tilde \psi(\tilde \phi) = (q + Q)
\psi(\phi, \Phi)$ such that the charge $\tilde q$  on the capacitor plate is
the conjugate variable to $\tilde \phi$. With that, we have decompactified the
phase $\phi$ (defined on the interval $[0,\Phi_Q]$) to $\tilde \phi$ (defined
on the complete real line). This gives the conventional form of the fluxonium
Hamiltonian (acting on the wavefunction $\tilde \psi$)
\begin{align}\label{eq:fluxonium_ham} \tilde H_\text{flux} = \frac{\tilde
q^2}{2C} - E_J \cos\biggl(\frac{2\pi
	\tilde \phi}{\Phi_Q}\biggr) + \frac{1}{2 L} (\tilde
  \phi-\Phi^\text{ex})^2,
\end{align}
with $\tilde q$ the conjugate variable to $\tilde \phi$. Alternatively, in the
path integral formulation, one can start with
\eqref{eq:fluxonium_ham_alternative} and integrate out the harmonic mode $Q$
in order to arrive at~\eqref{eq:fluxonium_ham}~\cite{apenko:89,zaikin:90}. 

It has previously been shown that in the limit $L \rightarrow \infty$,
selection rules emerge from the Hamiltonian~\eqref{eq:fluxonium_ham} which
limit the dynamics of the decompactified variable $\tilde \phi$ to the
dynamics of a compact variable $\phi$ corresponding to the system without a
shunt~\cite{koch:09}.  The origin of the selection rules is made transparent
by the Hamiltonian~\eqref{eq:fluxonium_ham_alternative} which shows that
polarization charge $Q$ becomes conserved in the limit $L \rightarrow \infty$.
The explicit separation of the transport of $q$ over the Josephson junction
and the flow of polarization charge $Q$ through the shunt in the
Hamiltonian~\eqref{eq:fluxonium_ham_alternative} clearly brings out the
different time scales associated with the two processes.  This fact makes it
very useful for the derivation of an effective fluxonium Hamiltonian as we
will discuss in Sec.~\ref{sec:fluxonium}.

As another example of the mixed formulation, we
discuss in the App.~\ref{app:mixed_formulation_circuit_vogt} the derivation of
a Hamiltonian for the experimental setup of Ref.~\onlinecite{vogt:15}.

\section{Applications}\label{sec:applications}
As we have discussed in the previous sections, Josephson junctions cannot be
handled directly using loop charges. On the other hand, it is well-known that
Josephson junctions are approximately self-dual~\cite{mooij:06} and can
behave as nonlinear capacitors at low energies. As we now want to show, this
yields an approximate way of incorporating Josephson junctions in the loop
charge description.

In particular, we discuss the example of a single Josephson junction:  the
effective nonlinear capacitor is given by the $2e$-periodic ground-state
energy $\varepsilon_0(Q)$, where $Q$ is the polarization charge. An
instructive way to understand the $2e$ periodicity is provided by writing the
total charge on capacitor plate as the sum $q+Q$ of the integer charge $n =
q/2e$ and the continuous polarization charge $Q$, cf.\
Eq.~\eqref{eq:fluxonium_ham}.\cite{schon:90} While the former corresponds to
(excess) Cooper-pairs on the island, the latter models the polarization
charge, i.e., continuous displacements of negative and positive charges on the
island against each other due to polarizing electric fields.  The unusual
aspect of the Josephson junction is the fact that it allows exchange of
individual Cooper-pairs through tunneling, whereas the polarization charge
remains fixed due to the insulating layer of the Josephson junction. As a
result, a Josephson junction is only able to screen the charges in units of
$2e$ yielding the periodic ground state energy  $\varepsilon_0(Q)$.

The separation of the charge $q+Q$ remains useful when shunting the Josephson
junction by a large (complex) impedance $Z_\omega$ that allows the exchange of
the polarization charge between the capacitor plates.  As long as the
impedance is large, there will be an adiabatic separation of the fast flow of
integer charges $n$ through the Josephson junction and the polarization charge
flow through the shunt. We will make this idea in two examples more explicit.

\subsection{Fluxonium}\label{sec:fluxonium}
We now want to apply this idea in the description of the fluxonium
circuit of Fig.~\ref{fig:fluxonium_circuit}.\cite{manucharyan:09}   In the
limit of large inductance $L$, the shunt impedance $Z_\omega=-i \omega L$
becomes large and we can perform the adiabatic decoupling of the polarization
charge $Q$ and the phase $\phi$ in the fluxonium
Hamiltonian~\eqref{eq:fluxonium_ham_alternative}.  To that end, we introduce
the (instantaneous) eigenstates $u_{Q,s}(\phi)$ of the Cooper-pair box
Hamiltonian
\begin{align}\label{eq:cpb_hamiltonian}
H_\text{cpb} = \frac{1}{2C}\biggl(-i\hbar\frac{\partial}{\partial \phi}
+Q\biggr)^2 
- E_J \cos\biggl(\frac{2\pi \phi}{\Phi_Q}\biggr), 
\end{align}
such that $H_\text{cpb}\, u_{Q,s}(\phi) = \varepsilon_s(Q) \, u_{Q,s}(\phi)$
holds, where $\varepsilon_s(Q)$ is the $2e$-periodic instantaneous eigenenergy
to the (constant) polarization charge $Q$. In the adiabatic approximation, we
make the ansatz $\psi(\phi, Q) =  u_{Q,s}(\phi) \chi_s(Q)$ for the total
wavefunction of $H_\text{flux}$ in Eq.~\eqref{eq:fluxonium_ham_alternative}.
Inserting this ansatz and neglecting derivatives of $u_{Q,s}$ with respect to
$Q$, we arrive at the lowest-order adiabatic
approximation~\cite{bohm,zorin:06,koch:09}
\begin{align}\label{eq:fluxonium_ham_phaseslip}
H_s = \frac{1}{2L}\biggl( i \hbar \frac{\partial}{\partial Q}
 - \Phi^\text{ex} \biggr)^2 + \varepsilon_s(Q)
\end{align}
for the Hamiltonian of the wavefunction $\chi_s(Q)$ which is $2e$ periodic.
The Hamiltonian \eqref{eq:fluxonium_ham_phaseslip} is the (passive) dual
description a Josephson junction shunted by a large impedance as alluded to in
the introduction and depicted in Fig.~\ref{fig:fluxonium_circuit}.

We want to comment on the connection of the wavefunctions $\chi_{s,n}(Q)$ for
the $n$-th eigenstate obtained in this manner to the wavefunction $\tilde
\psi(\tilde \phi)$ of the (conventional) fluxonium Hamiltonian $\tilde
H_\text{flux}$ of Eq.~\eqref{eq:fluxonium_ham}. Using the relation
\eqref{eq:bloch_theorem} as well as the adiabatic ansatz, we obtain
\begin{align}\label{eq:fluxonium_approximate_eigenstates}
\tilde\psi_{s,n}(\tilde \phi) 
= \int_{0}^{2e}\! \frac{d Q}{2\pi \hbar} 
 u_{Q,s}(\tilde \phi) \chi_{s,n}(Q) e^{i Q \tilde \phi/\hbar}, 
\end{align}
as an approximate expression of the exact eigenstates.

\begin{figure}[tb]
\includegraphics[width=\columnwidth]{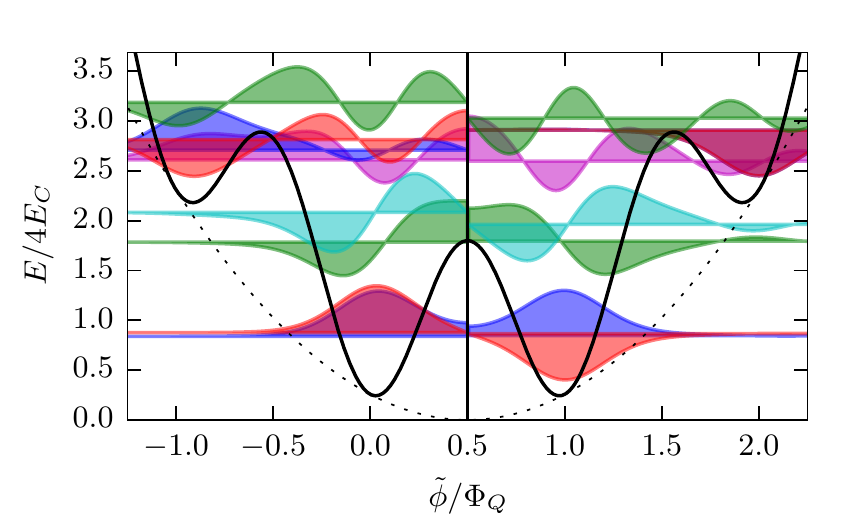}%
\caption{Exact ($\tilde \phi < \Phi_Q/2$) and approximate ($\tilde \phi >
  \Phi_Q/2$)
  fluxonium wave functions for $\Phi^\text{ex}=\Phi_Q/2$.  The exact wave
  functions are computed by exact diagonalization of the full
  Hamiltonian~\eqref{eq:fluxonium_ham} and the approximate wave functions are
  obtained by computing eigenstates $\chi_{s,n} (Q)$ of the adiabatic
	Hamiltonian~\eqref{eq:fluxonium_ham_phaseslip} and using
	formula~\eqref{eq:fluxonium_approximate_eigenstates}. The states live
	in a potential (solid black line) 
	composed of a harmonic contribution (dashed black line) due to the
	inducance with an inductive energy $E_L = (\Phi_Q/2\pi)^2/L$ 
	and the superposed cosine potential due to the Josephson junction with
	the Josephson energy $E_J$. The parameters
	$E_J/4 E_C = 0.9$ and $E_L/4 E_C = 0.052$ (with the capacitive energy $E_C = e^2/2C$)
	correspond to the qubit discussed in Ref.~\onlinecite{manucharyan:09}.
	The wave functions of both Hamiltonians can be chosen real due to the
	symmetry under $(\phi, \Phi) \mapsto(-\phi,\Phi_Q - \Phi)$ and are centered
	vertically at their corresponding energy eigenvalue. }%
\label{fig:fluxonium_spectra} %
\end{figure}

\begin{figure}[tb]
\includegraphics[width=\columnwidth]{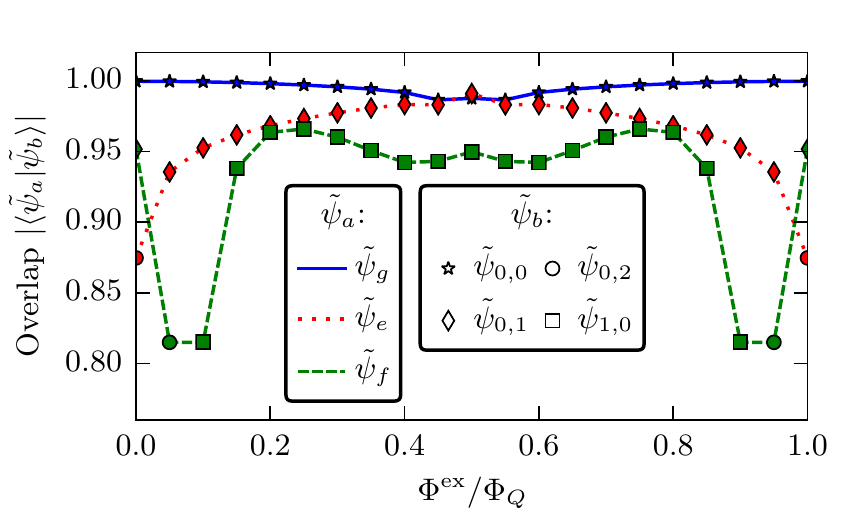}%
\caption{Approximate fluxonium eigenstates $\tilde\psi_{s,n}$ which have the maximum
overlap with either of the three lowest energy states $\tilde\psi_g$, $\tilde\psi_e$,
$\tilde\psi_f$ of the exact fluxonium Hamiltonian~\eqref{eq:fluxonium_ham}.  The
approximate eigenstates are calculated via
Eq.~\eqref{eq:fluxonium_approximate_eigenstates}, using the eigenstates of the
Hamiltonian~\eqref{eq:fluxonium_ham_phaseslip} which arises from projection on
band $s$ of the Cooper-pair box Hamiltonian~\eqref{eq:cpb_hamiltonian}. }%
\label{fig:fluxonium_overlaps} %
\end{figure}

To highlight the accuracy of the approximate expression
\eqref{eq:fluxonium_approximate_eigenstates}, we have numerically calculated
the eigenstates $\tilde \psi_m(\tilde \phi)$ of the
Hamiltonian~\eqref{eq:fluxonium_ham}, as explained in App.~\ref{sec:numerov},
and the eigenstates $\chi_{s,n}(Q)$ and $u_{Q,s}(\phi)$ of the
Hamiltonians~\eqref{eq:fluxonium_ham_phaseslip} and
\eqref{eq:cpb_hamiltonian}.  In Fig.~\ref{fig:fluxonium_spectra}, we show the
comparison of the exact eigenstates to the approximate
eigenstates~\eqref{eq:fluxonium_approximate_eigenstates} for $\Phi^\text{ex} =
\Phi_Q/2$.  Note that the wave functions can be chosen real due to the
symmetry under $\tilde \phi \mapsto \Phi_Q- \tilde \phi$ (or, $(\phi, \Phi)
\mapsto(-\phi,\Phi_Q - \Phi)$, respectively) and are centered vertically at
their corresponding energy value.  Especially for the low-lying states, one
sees good agreement between the exact eigenstates and the approximate
states~\eqref{eq:fluxonium_approximate_eigenstates}.  In particular, for the
exact lowest energy states $\tilde \psi_g$, $\tilde\psi_e$ of the fluxonium
Hamiltonian~\eqref{eq:fluxonium_ham}, we find the correspondence
\begin{align}
\bigl(\tilde \psi_g, \tilde \psi_e \bigr)
\mapsto (\tilde \psi_{0,0}, \tilde \psi_{0,1}),
\end{align}
i.e., the states $\tilde \psi_g$, $\tilde \psi_e$ are all associated with the
lowest $s=0$ band of the Cooper-pair box.  As we show in
Fig.~\ref{fig:fluxonium_overlaps}, this property persists for the entire range
of external flux $\Phi^\text{ex}$.  In particular around the experimentally
relevant flux bias of half a flux quantum, $\Phi^\text{ex} = \Phi_Q/2$, we
find overlaps $|\langle \tilde\psi_g | \tilde\psi_{0,0}\rangle|$, $|\langle
\tilde\psi_e | \tilde\psi_{0,1}\rangle|$ well above $0.95$.  We thus arrive at
the conclusion that the fluxonium can effectively be understood as a
phase-slip junction with a constitutive relation $V = f_V(Q)= \epsilon_0'(Q)$.
Instead of using the original fluxonium circuit from
Fig.~\ref{fig:fluxonium_circuit}(a), we may therefore obtain an accurate
description in terms of the simpler circuit depicted in
Fig.~\ref{fig:fluxonium_circuit}(b), which follows from the first circuit by
replacing the Josephson junction and its associated capacitance by a phase
slip junction.

The circuit from Fig.~\ref{fig:fluxonium_circuit}(b) yields a simplified
fluxonium description which may, e.g., be convenient in order to
understand the effects of environmental noise.  As an example, we consider the
case of a noisy inductor which we model by an additional resistor $R$ in
series with the inductance $L$.  The flux $\phi$ over the resistor couples
linearly to the current $\dot Q$ and we can therefore apply the results of
Sec.~\ref{sec:dissipation}.  Using standard results for
qubits~\cite{makhlin:01, nazarov_noise_schoelkopf}, one arrives at a relaxation rate
\begin{align}
\Gamma_1 = \frac{|\langle \chi_{0,0} |  \partial_Q | \chi_{0,1}
	\rangle|^2}{L^2}\, 
S_\phi(E_{01}/\hbar),
\end{align}
where $E_{01}>0$ denotes the energy difference between the states $\chi_{0,1}$
and $\chi_{0,0}$ and $S_\phi(\omega) = \int\! d t \, e^{i\omega t} \langle
\phi(t) \phi(0)\rangle = 2\hbar R (n_\omega+1)/\omega$ is the spectral density
of flux fluctuations over the resistor. In units of the RL-time
$\tau_\text{RL} = L/R$, the result reads $\Gamma_1 \tau_\text{RL} = (n_B+1)
\Phi_{01}^2 /L E_{01} $ with $n_B$ the photon number at frequency $\omega
=E_{01}/\hbar$. As a result, the decay $\Gamma_1$ is proportional to  the
ratio of the magnitude of energy fluctuations $\Phi^2_{01}/L$ due to the
(quantum) fluctuations of $\Phi$ to the energy difference of the
transition.

\subsection{$0$-$\pi$ qubit}
\label{sec:zeropi}
As another example, we consider the $0$-$\pi$ qubit, which is based on a
special type of Josephson inductance that is $\Phi_Q/2$-periodic in the phase
$\phi$.  This has to be contrasted with the $\Phi_Q$-periodicity found in
conventional Josephson junctions.  There exist two different proposals for its
realizations.  The first proposal, the superconducting current mirror, is
based on an energetic suppression of single Cooper-pair
tunneling~\cite{kitaev:06a}, whereas the second proposal, the Josephson
rhombus, is based on destructive interference of single Cooper-pair tunneling
guaranteed through symmetry~\cite{doucot:02,doucot:12}.  Independent of the
specific way the $\Phi_Q/2$-periodic junction is realized, its effective
Hamiltonian can be written as
\begin{align}\label{eq:zeropi_ham}
H = 4 E_C (q+ Q)^2  - E_{J2} \cos(4 \pi \phi/\Phi_Q),
\end{align}
where $q = -i \hbar \partial/\partial \phi$ is conjugate to $\phi$, $E_{J2}$
gives the strength of the $\Phi_Q/2$-periodic junction and we have included a
charging energy with polarization charge $Q$.

\begin{figure}[tb]
\includegraphics[width=\columnwidth]{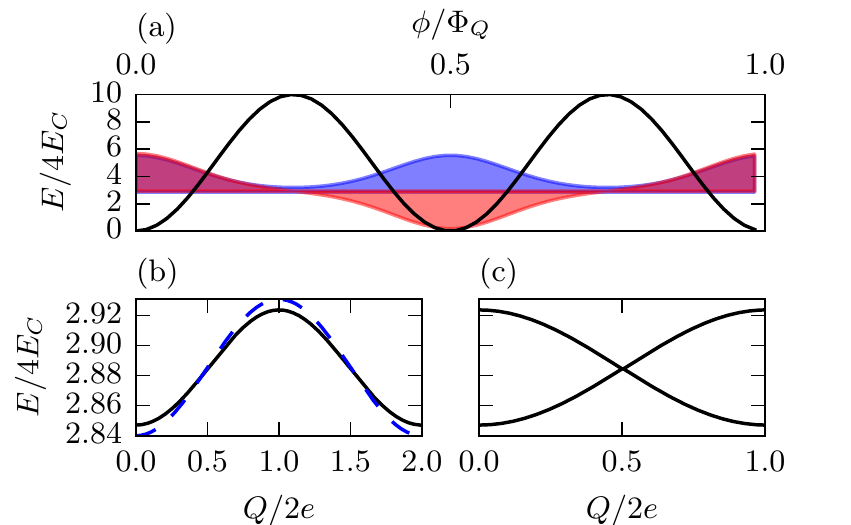}%
\caption{In (a), we show the two lowest-energy wave functions of the
$0$-$\pi$ Hamiltonian~\eqref{eq:zeropi_ham} for $E_{J2} = 20 E_C$ and $Q = 0$.
The wave functions can be chosen real and are centered vertically at their
corresponding energy.  One observes the even or odd character of the
eigenstates under translations by $\Phi_Q/2$ which reflects the Cooper-pair
parity of the states.  In (b), we illustrate the Bloch bands originating
from the choice of a $\Phi_Q/2$-periodic unit cell, resulting in Bloch-band
periodicity of $4e$.  The asympotic estimate~\eqref{eq:zeropi_dispersion}
valid for $E_{J2} \gg E_C$ is shown as a blue dashed line.  In (c), we
illustrate the folded zone scheme corresponding to the choice of a
$\Phi_Q$-periodic unit cell which arises from (b) by folding the part of the
Bloch bands for $Q > 2e$ back.  The two resulting bands differ
in Cooper-pair parity.  The band crossings at $Q = e$ are protected as long as
Cooper-pair parity is conserved.}%
\label{fig:zeropi_bands} %
\end{figure}

There exist two possible choices of qubit states.  When the junction strength
is much larger than the charging energy, $E_{J2}/E_C \gg 1$, the states can be
approximated as states localized at the potential minima $\phi = 0$ or $\phi =
\Phi_Q/2$ of the junction term.  On the other hand, it is clear that the
correct eigenstates of the Hamiltonian~\eqref{eq:zeropi_ham} are characterized
by Cooper-pair parity as a good quantum number, since the $E_{J2}$ term only
connects charge states differing by $4e$.  Indeed, tunneling between the
minima of the junction leads to a hybridization of the states localized at
$\phi = 0$ or $\phi = \Phi_Q/2$ into odd and even superpositions
$\psi_o$,
$\psi_e$ which are in direct correspondence to states characterized by odd
or even Cooper-pair parity~\cite{bell:14}.  This is illustrated in
Fig.~\ref{fig:zeropi_bands}(a) for $E_{J2} = 20 E_C$ and $Q=0$.  Going over
to a Bloch band description with the choice of a $\Phi_Q/2$-periodic unit cell
allows mapping the Hamiltonian~\eqref{eq:zeropi_ham} to the Hamiltonian of the
conventional Cooper-pair box.  One can then use the semiclassical results for
the $2e$-periodic charge dispersion of the lowest band of the conventional
Cooper-pair box~\cite{koch:07}.  After the appropriate scaling, it maps to the
$4e$-periodic charge dispersion $\varepsilon_0 = -A \cos(\pi Q/2e) +
\text{const.}$ with bandwidth $2A$, where $A$ is given by
\begin{align}\label{eq:zeropi_dispersion}
A = 2^{6} \sqrt{\frac{2}{\pi}}
\Biggl(\frac{E_{J2}}{8 E_C}\Biggr)^{\tfrac 3 4} 
E_C e^{-\sqrt{2 E_{J2}/E_C}}.
\end{align}
For the lowest band, the exact charge dispersion (solid line) and its
asymptotic expression~\eqref{eq:zeropi_dispersion} (dashed line) is
illustrated in Fig.~\ref{fig:zeropi_bands}(b) for the same parameters
$E_{J2} = 20 E_C$ as in (a).  Going back to a $\Phi_Q$-periodic unit cell
corresponds to folding the Bloch-bands for $Q > 2e$ back to the origin.  The
resulting band structure is displayed in Fig.~\ref{fig:zeropi_bands}($c$).
The states from the lowest two bands are the qubit states $\psi_e$,
$\psi_o$ corresponding to even or odd Cooper-pair parity.  In the regime
$E_{J2} \gg E_C$, the gap $E_{eo}$ between the two states roughly scales as
$E_{eo} = 2A \propto e^{-\sqrt{2E_{J2}/E_C}}$.

The question of which choice of states adequately describes the qubit depends
on the size of perturbations that yield transitions between states of
different Cooper-pair parity.  Such a perturbation is, e.g., a finite
amplitude $E_{J1}$ for tunneling of conventional Cooper-pairs.  An amplitude
$E_{J1}$ that is much larger than the gap $E_{eo}$ will lead to a rapid
dephasing of the superpositions in the states $\psi_e$, $\psi_o$ and
effectively project back to the states localized at the potential minima.

For the following, we are interested in the regime where $E_{J1}$ is smaller
than $E_{eo}$.  Note that this is, e.g., the regime of the experiments
discussed in Ref.~\onlinecite{bell:14}.  In this case, the Cooper-pair parity
and the offset charge $Q$ in the interval $(0,2e)$ remain good quantum numbers
and the level structure can be represented as indicated in
Fig.~\ref{fig:zeropi_bands}(c).  Note that the crossing of the two level
curves is protected as long as Cooper-pair parity is conserved.

It is intriguing to note that there is an obvious duality between the charge
dispersion of the $0$-$\pi$ qubit shown in Fig.~\ref{fig:zeropi_bands}(c)
and the flux dispersion of a junction connecting two Majorana bound states
with energy (fractional Josephson effect)~\cite{kitaev:01}
\begin{align}\label{eq:majorana_junction}
H = \pm \cos(\pi \phi/\Phi_Q),
\end{align}
where the choice of the plus or minus sign is related to the occupation parity
of the nonlocal fermion hosted by the Majorana bound states.  Dual to the
treatment of the $0$-$\pi$ qubit, one can describe the $2\Phi_Q$-periodic
Majorana junction in terms of a folded zone-scheme in a $\Phi_Q$-periodic unit
cell, leading to a similar picture as in Fig.~\ref{fig:zeropi_bands}(c) but
with $Q/2e$ replaced by $\phi/\Phi_Q$.  Now the two bands differ in
superconducting flux quantum parity and the crossing at $\Phi_Q/2$ is
protected as long as flux quantum parity is preserved.  This corresponds to an
absence of conventional Josephson junctions in a loop with the Majorana
junction through which conventional $\Phi_Q$ phase-slips may
occur~\cite{heck:11}.

Embedding the $0$-$\pi$ circuit in a large-impedance environment as discussed
in Sec.~\ref{sec:applications} leads to a low-energy description by
states living in the charge-dispersion bands from
Fig.~\ref{fig:zeropi_bands}(c).  With this starting point, one may consider
more complex circuits.  We thus arrive at there intriguing conclusion that the
$0$-$\pi$ qubit may allow us to explore the plethora of proposals existing for
Majorana qubits~\cite{beenakker:13, stanescu:13} from a dual perspective.

\section{Conclusions}
In this paper, we have discussed a charge-based approach to circuit
quantization using loop charges which are the time-integrated currents
circulating in the loops of a planar circuit.  We have shown that the
appropriate circuit Lagrangian can be read off the electrical network using a
set of simple rules. In this approach, we obtain a local Hamiltonian
description in terms of charges in a planar circuits of arbitrary topology.
We have discussed how to handle dissipative elements by going over from closed
systems to open systems.

We have shown explicitly that a passive duality transformation relates the
charge-based circuit description in terms of loop charges to the flux-based
description in terms of node fluxes which is conventionally employed for the
quantization of superconducting circuits.  While the flux-based formulation is
convenient for the description of charge currents, the charge-based
formulation yields a simple description whenever the dynamics is characterized
by flux currents.  In particular, we have argued that passive duality
transformations are useful for Josephson junctions in large-impedance
environments, which behave as nonlinear capacitors supporting a quantized flux
flow at low energies.

We have shown that the loop charge formulation can be used more generally for
the description of arbitrary circuits involving phase-slip junctions which are
nonlinear capacitors electromagnetically dual to Josephson junctions.  We have
explained that electromagnetic duality can be used as an active transformation
yielding new circuits whose charge dynamics is identical to the flux dynamics
of the original circuit.  We have shown how the loop charge formalism allows
the straightforward construction of such active duality transformations. In
particular, Josephson junctions have to be replaced by phase-slip junctions.
The duality between the node fluxes and the loop charges guarantees that the
loop charges are useful for the  description of latter circuits in the same
way that node fluxes are useful for Josephson junction circuits.

We have introduced a mixed circuit description in terms of loop charges and
node fluxes.  We have shown that the mixed formulation gives additional
insights into the decompactification of the flux $\phi$ over a Josephson
junction that is shunted by an inductor.

We have explicitly  illustrated how passive duality transformations yield
simplified circuit descriptions for Josephson junctions shunted by large
impedances using the fluxonium qubit and the $0$-$\pi$ qubit as an example.
We have shown that regarding the fluxonium as a nonlinear capacitor yields an
approximate though accurate description of the qubit states for relevant qubit
parameters.  We have illustrated how this may be used, e.g., for a simplified
description of relaxation caused by environmental noise.  As another example,
we have considered the $0$-$\pi$ qubit.   We have shown that in the absence of
conventional Cooper-pair tunneling, the junction dynamics becomes
electromagnetically dual to the dynamics of a Majorana Josephson junction.

From this work, several interesting routes arise that could be explored in
the future.  It will be highly interesting to use the loop charge formalism
for quantitative analysis of recent experiments involving phase-slip junctions.
It will also be interesting to exploit the duality of the $0$-$\pi$ qubit to a
Majorana junction and explore existing proposals for Majorana physics from a
dual perspective.

\begin{acknowledgments} 
The authors would like to acknowledge helpful
discussions with David DiVincenzo and Nikolas Breuckmann.  The authors are
grateful for support from the Alexander von Humboldt foundation.
\end{acknowledgments}

\appendix

\section{Mathematical preliminaries}
\label{app:math}
For the convenience of the reader, we here want to rederive the standard
result of circuit analysis~\cite{chen,thulasiraman} that in a planar circuit,
loop charges $\vec Q$ determine all the branch currents $\dotvec{q}^\text{br}$
in such a way that the Kirchhoff current law is fulfilled.  Along the way, we
will recall a few standard mathematical results about graphs that will be used
in the remainder of the appendix.  More information can be found in the
literature~\cite{chen,thulasiraman}.

We first need to show that there is an independent current for each of the $m$
chords of the spanning tree.  To see that the Kirchhoff current law implies
precisely $m$ independent currents, we make use of the basis node-edge
incidence matrix $A$, which is a $\mathbb{R}^{(n-1) \times b}$ matrix for the
$n-1$ nodes without the ground node and the $b$ branches.  Its entries $A_{ij}
\in \{1,-1\}$ indicate whether branch $j$ enters ($-1$) or leaves $(+1)$ node
$i$.  Given a vector $\vec{q}^\text{br}$ of branch charges, the Kirchhoff
current law can be expressed as $A \dotvec{q}^\text{br} = 0$.  A decomposition
of $\vec q^\text{br} = (\vec q^\text{ch}, \vec q^\text{tr})$ into the vector of
chord charges $\vec q^\text{ch}$ and tree charges $\vec q^\text{br}$ gives rise
to a corresponding decomposition of $A = (A_\text{ch}, A_\text{tr})$.  Since
there are no loops in a tree, we have the result $A_\text{tr} \vec v \neq 0$
for every vector $\vec v \in \mathbb{R}^b$, implying that $A_\text{tr}$ has
full rank and the inverse of $A_\text{tr}^{-1}$ is well-defined~\cite{*[{}][{,
		Theorem 2.2.}] chen}.  One can also show the result $\left|\det
	A_\text{tr}\right| = 1$.  Using that $A_\text{tr}^{-1}$ is invertible,
we obtain the relation $\dotvec{q}^\text{tr} = -A_\text{tr}^{-1} A_\text{ch}
\dotvec{q}^\text{ch}$, showing that the $m$ chord charges $\vec q^\text{ch}$
fully specify all currents in the circuit.

Our intuitive notion that the loop currents give the correct number of
independent currents in a planar graph is confirmed by Euler's theorem for
connected planar graphs which is the relation $n - b + f = 2$, where $f$ is
the number of faces of a graph.  Using $b = m + n-1$, we obtain $f = m + 1$,
where the $+1$ arises since $f$ also counts the exterior of the planar graph
as a face.  This shows that the loop charges in the faces of the graph indeed
give the correct number of independent currents for a planar circuit.  More
generally, one can show~\cite{thulasiraman} that this is no longer case for a
nonplanar graph.

It remains to relate the chord charges $\vec{q}^\text{ch}$ more explicitly to
the loop charges $\vec Q$.  To characterize the change of variables from $\vec
q^\text{ch}$ to $\vec Q$, we note that we may characterize the loops of a
circuit in terms of the fundamental circuit matrix $B \in \mathbb{R}^{m \times
b}$, where each entry $B_{ij} \in \{1,-1\}$ indicates that the branch $j$ is
oriented in the same direction ($1$) or opposite ($-1$) to the arbitrarily
chosen orientation of the loop $i$ formed by the $i$-th chord and the branches
of the spanning tree.  The matrix $B$ obeys the important relation $A B^T = 0$
which expresses the fact that for each node that is part of some loop,
branches having the same incidence orientation with respect to the node will
necessarily have opposite orientations with respect to the loop.  From the
relation $A B^T = 0$ and the decomposition $A = (A_\text{ch}, A_\text{tr})$,
we obtain the expression $B' = (1, -A_\text{ch}^T (A_\text{tr}^{-1})^T)$ for
the fundamental circuit matrix corresponding to the loop basis induced by the
chords.  For the loop basis corresponding to the loop charges we have the more
general form $B = (B_\text{ch}, B_\text{tr})$ where $B_\text{ch}$ is
invertible since it is related to the identity matrix via a basis
transformation in loop space.  This finally gives the relation $\vec
q^\text{ch} = B_\text{ch}^T \vec Q$.

By definition of the matrices $A$ and $B$, we obtain the results
$\dotvec{q}^\text{br} = B^T \dotvec{Q}$ and $\dotvec{\phi}^\text{br} = A^T
\dotvec{\phi}$.  Making use of the relation $A B^T = 0$ shows that the branch
fluxes and branch charges defined in this way automatically fulfill the
Kirchhoff voltage law $B \dotvec{\phi}^\text{br} = 0$ and the Kirchhoff
current law $A \dotvec{q}^\text{br} = 0$.

\section{Duality in the path integral}
\label{app:duality}

Our starting point is expression~\eqref{eq:legendre_fourier_transform},
\begin{align}\label{eq:legendre_fourier_transform_app}
e^{(i/\hbar) \int^t\!dt'\, \mathcal{L}(\vec \phi^\text{br})} 
= \int \mathcal{D}[\vec q^\text{br}(t)] 
e^{(i/\hbar) \int^t\!dt' [ 
	\tilde{\mathcal{L}}(\vec{q}^\text{br})
	- \vec{q}^\text{br}\cdot \dotvec{\phi}^\text{br}]}. 
\end{align}
For the decomposition $\vec q^\text{br} = (\vec q^\text{ch}, \vec
q^\text{tr})$ of the branch charges, we have found in App.~\ref{app:math} the
relation $\dotvec{q}^\text{tr} = -A_\text{tr}^{-1} A_\text{ch}
\dotvec{q}^\text{ch}$, which shows that the chord charges
$\vec{q}^\text{ch}$ determine the tree charges $\vec{q}^\text{ch}$ up to
constant offset charges $\vec \lambda$.  We can make this explicit by
introducing the factor
\begin{align}
1 = \int \mathcal{D}[\vec \lambda(t)] \, 
 \delta[(\vec q^\text{tr} + A_\text{tr}^{-1} A_\text{ch} \vec
 q^\text{ch} - \vec \lambda)(t)]
\end{align}
into the integral~\eqref{eq:legendre_fourier_transform_app}.  Using the
relation $\vec \phi^\text{br} = A^T \vec \phi$ for the vector of node fluxes
$\vec \theta$ and performing the integration over the tree charges $\vec
q^{\text{tr}}$ yields
\begin{multline}
e^{(i/\hbar) \int^t\!dt'\, \mathcal{L}(\vec \phi^\text{br})} 
= \int \mathcal{D}[\vec \lambda(t)] \int \mathcal{D}[\vec q^\text{ch}(t)] 
\\
\times 
e^{(i/\hbar) \int^t\!dt' [\tilde{\mathcal L}(\vec q^\text{ch}, -A_\text{tr}^{-1}
	A_\text{ch} \vec{q}^\text{ch}+\vec \lambda) - \vec \lambda
	A_\text{tr}^T \dotvec{\phi}]}.
\end{multline}
Performing a partial integration on the term $-i \vec \lambda A_{tr}^T
\dotvec{\phi}/\hbar$ in the exponent, inserting the resulting expression in
Eq.~\eqref{eq:time_evolution_pathint} and performing the integration over the
node fluxes $\vec{\phi}$, we obtain a constraint at each point in time in
terms of the delta function  $\delta[A_\text{tr} \dotvec \lambda(t) ]$.  Since
$A_\text{tr}$ has full rank and obeys $|\det A_\text{tr}| = 1$, this is
equivalent to demanding $\dotvec{\lambda} = 0$ for all times. We resolve this
constraint by demanding that offset charges are constant, $\bm \lambda (t)
\equiv \bm \lambda$. In fact the value of $\bm \lambda =0 $ is fixed by the
boundary condition that all the elements are uncharged for $t\to-\infty$. We
thus obtain
the representation
\begin{multline}
e^{-i H t/\hbar}
\rightarrow 
\int \mathcal{D}[\vec q^\text{ch}(t)] \\
\times
e^{(i/\hbar) \int^t\!dt'\, 
  \tilde{\mathcal L}[\vec q^\text{ch}, -A_\text{tr}^{-1}
  A_\text{ch} \vec{q}^\text{ch}]},
\end{multline}
for the time-evolution operator. In a planar circuit, we may finally exploit
the relation $\vec q^\text{ch} = B_\text{ch}^T \vec Q$ and replace the
integration over $\vec q^\text{ch}$ by an integration over the loop charges
$\vec Q$.  We then recover expression~\eqref{eq:dual_path_integral} from the
main text.

\section{Equivalence of terms manifestly guaranteeing the Kirchhoff current
	law or the voltage law in the mixed formulation}
\label{app:mixed_formulation_current_conservation}
We want to prove the equality (up to a total time-derivative) of the term
$-\sum_i \phi^\partial_i \sum_j \dot q_{ij}$ manifestly guaranteeing the
Kirchhoff current law and the term $-\sum_i Q^\partial_i \sum_j \dot \phi_{ij}$
manifestly guaranteeing the Kirchhoff voltage law.

Let $P \in \mathbb{R}^{b \times b}$ be the matrix projecting on the branches
(the subgraph) that shall be described in terms of loop charges.  We note that
we have the identity
\begin{align}\label{eq:mixed_terms_current_law}
-\sum_i \phi^\partial_i \sum_j \dot q_{ij}
= - \vec \phi A P B^T \dotvec{Q},
\end{align}
where $B$ is the fundamental circuit matrix introduced in App.~\ref{app:math}
corresponding to the loop charges $\vec Q$.  This identity can be understood
by noting that $P B^T \dotvec{Q}$ is the projection of the vector of branch
currents onto the branches of the subgraph.  The expression $(A P B^T
\dotvec{Q})_i$ gives the current balance for each node $i$ of the subgraph.
According to the definition of the basis node-edge incidence matrix $A$, positive
currents flowing away from node $i$ come with a plus sign, while positive
currents flowing into node $i$ come with a minus sign.  In line with the
definition of the $\dot q_{ij}$, one thus obtains in both cases the current
flowing away from node $i$.  Crucially, due to the usage of the loop charge, the
current balance is nonzero only for the boundary nodes $i$ with corresponding
node flux $\phi^\partial_i$, which proves the equality.  Using the
orthogonality $A B^T = 0$ and performing a partial integration, we can rewrite
the expression~\eqref{eq:mixed_terms_current_law} as
\begin{align}\label{eq:mixed_terms_voltage_law}
- \vec \phi A P B^T \dotvec{Q}
&= -\vec Q B (1-P) A^T \dotvec{\phi}+ \text{(ttd.)} \nonumber \\
&= - \sum_i Q^\partial_i \sum_j \dot \phi_{ij} + \text{(ttd.)} 
\end{align}
where (ttd.) stands for a total time-derivative.  Here, $(1-P) A^T
\dotvec{\phi}$ is the vector of voltage drops over the branches of the
subgraph complement.  The expression $[B (1-P) A^T \dotvec{\phi}]_i$
gives the voltage balance for each loop $i$ of the subgraph complement, which
is nonzero only for the boundary loops $i$ with corresponding loop charges
$Q^\partial_i$.  This proves the last equality sign.

\begin{figure}[tb]
\centering
\includegraphics{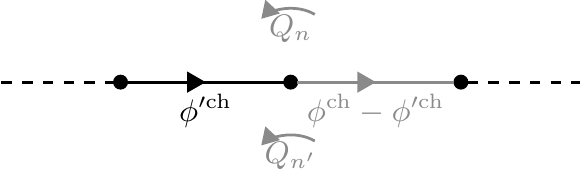}
\caption{In order to describe the presence of external fluxes, each chord
	branch of the circuit (solid lines) is split into two branches, one
	representing the original element (black solid line), the other
	representing the electromotive force (solid gray line).  As a
	consequence, the total flux $\phi^\text{ch}$ along the elements splits
	into the flux $\phi^{\prime
		\text{ch}}$ along the original element and the flux 
	$\phi^\text{ch} - \phi^{\prime \text{ch}}$ along the virtual branch.
	Ensuring the Kirchhoff laws therefore requires adding the terms
	$-(Q_n - Q_{n'})(\dot \phi^\text{ch} -  \dot \phi^{\prime \text{ch}}) =
	-q^\text{ch} (\dot \phi^\text{ch} - \dot \phi^{\prime \text{ch}})$ with
	the chord charge $q^\text{ch} = Q_{n} - Q_{n'}$ to the
	Lagrangian.\label{fig:external_fluxes}}
\end{figure}

\section{Proof of the rules for the inclusion of external
	fluxes using the mixed formulation}
\label{app:mixed_formulation_external_fluxes}
In this section, we want to show that the mixed formulation allows to
understand the origin of the rules for the inclusion of external fluxes into
the node flux formulation that were given in Ref.~\onlinecite{devoret:96}.

To that end, let us assume the presence of fluxes $\vec \Phi^\text{ex}$ in the
loops corresponding to the loop charges $\vec Q$.  We split each chord of the
circuit graph into two branches, one which represents the original chord
element and a second virtual branch which represents the electromotive force
due to the external flux.  As a consequence of the splitting, the total flux
$\vec \phi^\text{ch}$ over the chord and the virtual branch will split up into
a flux $\vec \phi^{\prime \text{ch}}$ over the chord element and a flux $\vec
\phi^\text{ch} - \vec \phi^{\prime \text{ch}}$ over the virtual branch.
Describing the virtual element in terms of charges requires adding the terms
$\dotvec{Q} \cdot \vec{\Phi}^\text{ex} - \vec q^\text{ch} \cdot
(\dotvec{\phi}^\text{ch} - \dotvec{\phi}^{\prime \text{ch}})$ to the
Lagrangian, c.f.~Fig.~\ref{fig:external_fluxes}.  As discussed in
App.~\ref{app:math}, the chord charges $\vec q^\text{ch}$ are related to the
loop charges $\vec Q$ according to $\vec q^\text{ch} = B_\text{ch}^T \vec Q$
with the invertible matrix $B_\text{ch}$.  Since the loop charges $\vec Q$ are
not dynamic, their equations of motion yield a constraint
$\dotvec{\phi}^{\prime \text{ch}} = \dotvec{\phi}^\text{ch} + B_\text{ch}^{-1}
\dotvec{\phi}^\text{ex}$. 

For a chord $b$ with an orientation that is consistent (inconsistent) with the
counter-clockwise orientation of its corresponding chord loop, the entries
$(B_\text{ch}^{-1})_{bl}$ are given by $+1$ ($-1$) for all loops that lie
within the face having the chord loop as its boundary and zero for all other
loops.  That means that all the non-zero entries in the rows of
$B^{-1}_\text{ch}$ are of absolute value $1$ and have the same sign.  To see
that this description of the entries yields indeed the inverse of
$B_\text{ch}$, let us consider the expression
\begin{align}
M_{bb'} = \sum_{l} (B^{-1}_\text{ch})_{bl} (B_\text{ch})_{lb'}.
\end{align}
We need to show that $M_{bb'} = \delta_{b b'}$.  When $b \neq b'$, the chord
$b'$ lies either outside or inside the face having the chord loop
corresponding to $b$ as its boundary.  It cannot lie on the boundary of the
face, i.e., it cannot be a part of the chord loop corresponding to $b$, since
the chords uniquely specify a loop in the graph.  If it lies outside the face,
we obtain $M_{bb'} = 0$ by our characterization of the matrix
$B_\text{ch}^{-1}$.  If it lies inside the face, it forms part of two
neighboring loops $l$, $l'$ whose entries $(B_\text{ch})_{lb'}$,
$(B_\text{ch})_{l'b'}$ differ in sign.  Since the rows of $B_\text{ch}^{-1}$
all have the same sign we also obtain $M_{b b'} = 0$ upon summing over $l$.
For $b=b'$, there is only one loop $l$ which lies in the face having the chord
loop corresponding to $b$ as its boundary, and the entries
$(B_\text{ch}^{-1})_{bl}$, $(B_\text{ch})_{bl}$ are both either plus or minus
one, giving $M_{bb} = 1$.  Therefore, $M_{bb'} = \delta_{bb'}$.  This shows
that we may simply work with the original circuit graph without the virtual
branches, provided we add to each expression involving the flux in a chord the
external flux in its corresponding loop~\cite{devoret:96}.

\section{Additional example for the mixed formulation}
\label{app:mixed_formulation_circuit_vogt}
\begin{figure}[tbp]
\includegraphics[width=\columnwidth]{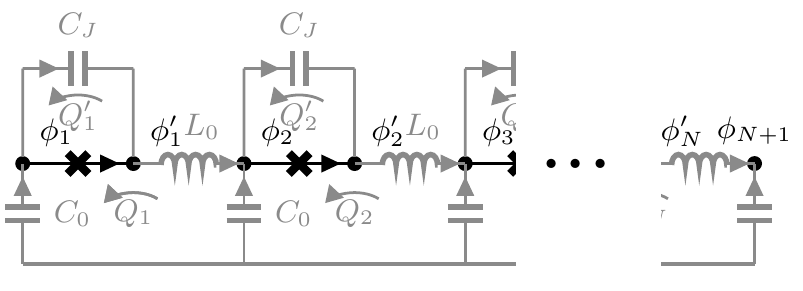}%
\caption{Circuit corresponding to the setup in Ref.~\onlinecite{vogt:15}.  We
	only want to describe the Josephson junction (the subgraph complement
	depicted in black) in terms of node fluxes, whereas we describe the
	rest of the circuit (the subgraph depicted in gray) in terms of loop charges.}%
\label{fig:circuit_vogt} %
\end{figure}

As an example, consider the circuit depicted in Fig.~\ref{fig:circuit_vogt}
which corresponds to the setup studied in Ref.~\onlinecite{vogt:15}. According
to the rules discussed in the main text, its Lagrangian reads 
\begin{align}
\mathcal{L} = \sum_{i=1}^N \Bigl[ \frac{1}{2 L_0} \dot Q_i^2 - \frac{1}{2 C_J} Q_i'^2 -
\frac{1}{2 C_0} (Q_i - Q_{i+1})^2  \nonumber \\
+ E_J \cos(2 \pi \varphi_i/\Phi_Q) - (Q_i' - Q_i) \dot \varphi_i \Bigr], 
\end{align}
where we have defined $Q_{N+1} = 0$ and $\varphi_i = \phi_i - \phi_{i'}$.
Note that the last term $-(Q_i' - Q_i) \dot \varphi_i$ just corresponds to the
term $-\sum_i Q^\partial_i \sum_j \dot \phi_{ij}$ that appears in the mixed
formulation as discussed in the main text.  Note that the term $\dot\varphi_i
Q_i$ enters with an overall plus sign since the voltage drop $\dot \varphi_i$
is measured in the direction opposite to the anticlockwise orientation of the
loop current $Q_i$.  There is no kinetic term for the coordinates $Q_i'$ such
that their Euler-Lagrange equations are algebraic with the solution $Q_i' = -
C_J \dot \varphi_i$.  Inserting this solution back into the Lagrangian and
performing the Legendre transformation with respect to $\dot \varphi_i$ and
$\dot Q_i$ yields the Hamiltonian
\begin{align}\label{eq:ham_vogt}
H = \sum_{i=1}^N \Bigl[ \frac{1}{2C_J}(q_i - Q_i)^2 - E_J \cos(2\pi
\varphi_i/\Phi_Q) \nonumber \\
+ \frac{1}{2C_0} (Q_i - Q_{i+1})^2 + \frac{1}{2 L_0} \Phi_i^2 \Bigr],
\end{align}
where $(q_i, \varphi_i)$ and ($\Phi_i, Q_i)$ are canonically conjugate pairs.
Eq.~\eqref{eq:ham_vogt} reproduces the result derived in
Ref.~\onlinecite{vogt:15}.

\section{Diagonalization of fluxonium using a higher-order matrix Numerov
	method}\label{sec:numerov}
An efficient way of diagonalizing the fluxonium Hamiltonian consists in
projecting the Hamiltonian onto the eigenstates of the harmonic part due to
charging energy and inductive shunt, and diagonalizing the resulting matrix.
The disadvantage of this method is the fact that it requires calculating
explicitly all matrix elements of the cosine potential using the harmonic
oscillator eigenstates.  This can be done analytically but the resulting
expressions are quite involved.  A more direct approach, which is simpler in
practice, consists in diagonalizing the Hamiltonian in real space. This
requires discretizing the second-order derivative operator.  For this, one
usually employs the lowest-order Numerov approximation of order
$\mathcal{O}(a^4)$, where $a$ is the lattice spacing.  The resulting
discretized Schroedinger equation can be recast in matrix
form~\cite{pillai:12} such that it can be conveniently solved by standard
(sparse) matrix methods.  It would seem natural to consider also higher-order
Numerov representations of the second-order derivative of order
$\mathcal{O}(a^{2r+2})$, where $r \in \mathbb{N}$, but they are normally
avoided due to stability issues~\cite{blatt:67}.  Interestingly, we have found
that stability is not a problem when solving the resulting eigenvalue problem
by standard (sparse) matrix methods instead of the conventional shooting
method; a method that will be described in the following.

We consider at time-independent Schroedinger equation of the form
\begin{align}\label{eq:numerov_eom} D^2 \psi(x) = \bigl[-i\partial_x +
A(x)\bigr]^2 \psi(x) = -f(x) \psi(x), \end{align}
where $D = -i\partial_x + A(x)$ is the covariant derivative operator and
$f(x)$ equals $f(x) = 2m [V(x)-E]$ for a Hamiltonian of the standard form $H =
(p + A)^2/2m + V(x)$.  For a wave function $\tilde \psi(x)$ defined as
\begin{align}
\tilde \psi(x) = e^{i \int^x dx' \, A(x')} \psi(x),
\end{align}
we find the relation
\begin{align}\label{eq:numerov_covariant_derivative}
e^{-i \int^x dx' \, A(x')} (-i \partial_x)^n \tilde \psi(x) = D^n \psi(x),
\end{align}
which gives a convenient way of evaluating the higher orders of the covariant
derivative acting on $\psi(x)$ through conventional derivatives of $\tilde
\psi(x)$.  Using Eq.~\eqref{eq:numerov_covariant_derivative}, we obtain
through Taylor expansion with respect to $\lambda$ the result
\begin{align}\label{eq:numerov_covariant_derivative_expr}
&e^{-i \int^x dx' \, A(x')}
\bigl[ \tilde \psi(x + \lambda) + \tilde \psi(x - \lambda)\bigr] \nonumber \\
&= \psi(x+\lambda) e^{i \int_x^{x+\lambda} dx' A(x')} + \psi(x-\lambda) e^{-i
	\int_{x-\lambda}^{x} dx' A(x')} \nonumber \\
&= \sum_{n=0}^\infty \frac{2 (-1)^n}{(2n)!} D^{2n} \psi(x) \lambda^{2n},	
\end{align}
which gives a relation between the values of the covariant derivatives $D^{2j}
\psi(x)$, $j \in \mathbb{N}_0$, and the value of the wave function $\psi(x)$
at positions $x\pm \lambda$.  Following ideas of Ref.~\onlinecite{tang:14}, we
stop the expansion~\eqref{eq:numerov_covariant_derivative_expr} at $n = r$ and
evaluate~\eqref{eq:numerov_covariant_derivative_expr} for values $\lambda = j
a$, $j \in \{-r,\dots,r\}\setminus\{0\}$, where $a$ is the lattice constant,
which gives $2r$ equations for the covariant derivative $D^{2j}\psi(x)$ and
the wave function values at points $\psi(x + j a)$.  Solving these equations
for $D^2 \psi(x)$ and $D^{2r}\psi(x)$ yields expansions of the form
\begin{align}
D^2 \psi(x) &= \frac{1}{a^2} \sum_{j=-r}^{j=r} c_j \psi_j +
\mathcal{O}(a^{2r}) \label{eq:numerov_D2}\\
D^{2r} \psi(x) &= \frac{1}{a^{2r}} \sum_{j=-r}^{j=r} d_j \psi_j +
\mathcal{O}(a^2),
\label{eq:numerov_D2r}
\end{align}
where we introduced the abbreviated notation $\psi_j = \psi(x+ja)$.  The
expansion coefficients $c_j$ and $d_j$ read
\begin{align}
c_j &= \sum_{k=1}^r \frac{2r^2 ((r-1)!)^2}{(r-k)!(r+k)!} \frac{(-1)^{k}}{k^2}
\biggl(-2 \delta_{j,0}  \nonumber \\
&\qquad \qquad\qquad \qquad\qquad
+\delta_{k,|j|} e^{i \int_x^{x+j a} dx'A(x')}\biggr),\\
d_j &= \frac{(-1)^{|j|} (2r)!}{(r-|j|)!(r+|j|)!} e^{i \int_x^{x+j a} dx'A(x')}.
\end{align}
Numerov's idea is to improve the accuracy of the expansion by a factor of
$a^2$ by exploiting the structure of the differential
equation~\eqref{eq:numerov_eom}.  Including the term of order $\lambda^{2n+2}$
in Eq.~\eqref{eq:numerov_covariant_derivative_expr} (that we previously
dropped in order to arrive at Eq.~\eqref{eq:numerov_D2}) and solving  for the
unknowns $D^{2j} \psi(x)$ with $j \in \{1,\dots,r\}$ while keeping
$D^{2r+2}\psi(x)$ as a free parameter yields
\begin{align}\label{eq:numerov_D2_improved}
D^2 \psi(x) &= \frac{1}{a^2} \sum_{j=-r}^r c_j \psi_j 
+ \frac{(r!)^2 a^{2r} D^{2r+2}\psi(x)}{(2r+1)! (r+1)}  \nonumber \\
& \qquad + \mathcal{O}(a^{2r+2}).
\end{align}
Acting on both sides of Eq.~\eqref{eq:numerov_eom} with $D^{2r}$ gives the
expression
\begin{align}\label{eq:numerov_trick}
D^{2r+2} \psi(x) = -D^{2r} \bigl[f(x) \psi(x)\bigr]
\end{align}
for $D^{2r+2}\psi(x)$.  Since we only need $D^{2r} [f(x) \psi(x)]$ to accuracy
$\mathcal{O}(a^2)$ in the expansion~\eqref{eq:numerov_D2_improved} of order
$\mathcal{O}(a^{2r+2})$, we can use the previously derived
expression~\eqref{eq:numerov_D2r}.  We then obtain the Numerov's expression
for the second-order covariant derivative
\begin{align}
D^2 \psi(x) = \frac{1}{a^2} \sum_{j=-r}^{r} c_j \psi_j 
- \frac{(r!)^2}{(2r+1)! (r+1)} \nonumber \\
\times \sum_{j=-r}^{r} d_j f_j \psi_j + \mathcal{O}(a^{2r+2}),
\end{align}
which is better by a factor of $a^2$ in accuracy compared to the naive
form~\eqref{eq:numerov_D2}.

Extending ideas of Ref.~\onlinecite{pillai:12}, we can convert this system of
equations into a generalized eigenvalue problem.  We introduce a matrix $A$
having $c_j/a^2$ on the $j$-th diagonal, where $j > 0$ refers to the upper
diagonals and $j < 0$ refers to the lower diagonals, a diagonal matrix $V =
\mathop{\rm diag}(V_j)$ representing the potential, and a matrix $B$ having
$-(r!)^2 d_j/(2r+1)!(r+1)$ on the $j$-th diagonal.  All of these matrices are
sparse and allow writing Eq.~\eqref{eq:numerov_eom} as the sparse generalized
eigenvalue problem
\begin{align}
	\biggl[\frac{1}{2m} A + (B+1)V\biggr]\vec \psi = E B
	\vec{\psi}, 
\end{align}
where $\vec \psi$ is the discretized wave function vector. This Hermitian
generalized eigenvalue problem can be solved efficiently by standard methods.

%\bibliography{/Users/jascha/bib/refs}

%\begin{thebibliography}{10}
%\makeatletter
%
%\bibitem{Note1}
%Remember that $q^\protect \text {br}_b$ counts the charge on one of the
%  capacitor plates while the capacitor itself is overall charge neutral.
%  Current flows may thus change $q^\protect \text {br}_b$ without violating
%  charge conservation.
%
%\bibitem{Note2}
%The presence of this loop can be rationalized by imagining an embedding of the
%  circuit on the surface of a sphere.

%\bibitem{Note3}
%In principle, we could have $\Phi = \phi + \Phi _0$ but $\Phi _0$ just
%  corresponds to a redefinition of the external flux $\Phi _\protect \text
%  {ex}$.
%\end{thebibliography}

\end{document}